\pdfoutput=1

\documentclass[sigconf,letterpaper]{acmart}
\graphicspath{ {./figures/} }
\usepackage[compact]{titlesec}
\usepackage{wasysym}
\usepackage{subcaption}
\usepackage{tabularx}
\usepackage{tabulary}
\usepackage{booktabs}
\usepackage{colortbl}
\usepackage{multirow}
\usepackage{color}
\usepackage{chngpage}
\definecolor{lightgray}{gray}{0.95}

\renewcommand{\paragraph}[1]{\vspace*{0.03in}\noindent\textbf{#1}}

\newcommand{\ie}{{\em i.e.}}
\newcommand{\eg}{{\em e.g.}}
\newcommand{\ea}{{\em et al.}}

\usepackage{microtype}
\usepackage[font=small,format=plain,labelfont=bf,textfont=it]{caption}

\title{{Inferring Streaming Video Quality from Encrypted Traffic: Practical Models and Deployment Experience}}
\author{Paul Schmitt$^\circ$,
Francesco Bronzino$^\ast$,
Sara Ayoubi$^\ast$,}
\author{
Guilherme Martins$^\circ$, Renata Teixeira$^\ast$,
and Nick Feamster$^\circ$
}

\affiliation{
  \institution{\hspace{-15mm}$^\circ$Princeton University\hspace{5mm}$^\ast$Inria, Paris}
}

\begin{document}

\newcommand{\sysname}{Network Microscope}

\setcopyright{none}
\settopmatter{printacmref=false} % Removes citation information below abstract
\renewcommand\footnotetextcopyrightpermission[1]{} % removes footnote with conference information in first column
\pagestyle{plain} % removes running headers

%\author{Paper \#XX - \pageref{p:lastpage} Pages}

\begin{abstract}
Inferring the quality of streaming video applications is important for Internet
service providers, but the fact that most video streams are encrypted makes it
difficult to do so. We develop models that infer quality metrics (\ie, startup delay and
resolution) for encrypted streaming video services. Our paper
builds on previous work, but extends it in several ways. First, the model works
in deployment settings where the video sessions and segments must be identified from a mix of
traffic and the time precision of the collected traffic statistics is more
coarse (\eg, due to aggregation). Second, we develop a single composite model 
that works for a range of different services (i.e., Netflix, YouTube, Amazon, and Twitch), as opposed to just a
single service. Third, unlike many previous models, the model performs
predictions at finer granularity (\eg, the precise startup delay instead of just detecting 
short versus long delays) allowing to draw better conclusions on the
ongoing streaming  quality. Fourth, we demonstrate the model is practical
through a 16-month deployment in 66 homes and provide new insights about the relationships
between Internet ``speed'' and the quality of the corresponding video streams,
for a variety of services; we find that higher speeds provide only minimal
improvements to startup delay and resolution.
\end{abstract}

\maketitle
\begin{sloppypar}

\section{Introduction}

Video streaming traffic is by far the dominant application traffic on today's
Internet, with some projections forecasting that video streaming will comprise
82\% of all Internet traffic in just three years~\cite{ciscoForecast}.
Optimizing video delivery depends on the ability to determine the quality of
the video stream that a user receives. In contrast to video content providers,
who have direct access to video quality from client software, Internet Service
Providers~(ISPs) must infer video quality from traffic as it passes
through the network. Unfortunately, because end-to-end encryption is
becoming more common, as a result of increased video streaming content over
HTTPS and QUIC~\cite{sandvine-encryption,MVI}, ISPs cannot directly observe
video quality metrics such as startup delay, and video
resolution from the video streaming protocol~\cite{gsma2015,dyer2015}.
The end-to-end encryption of the video streams thus presents ISPs with the
challenge of inferring video quality metrics solely from properties of the
network traffic that are directly observable.

Previous approaches infer the quality of a specific video service, typically
using offline modeling and prediction that is based on an offline trace in a
controlled laboratory setting~\cite{infocomPaper, dimopoulos2016measuring,
krishnamoorthi2017buffest}.  Unfortunately, these models are often not
directly applicable in practice because practical deployments (1)~have other
traffic besides the video streams themselves, creating the need to identify
video services and sessions; (2)~have multiple video services, as opposed to
just a single video service. Transferring the existing models to practice
turns out  to introduce new challenges due to these factors.

First, inference models must take into account the fact that real network
traffic traces have a mix of traffic, often gathered at coarse temporal
granularities due to monitoring constraints in production networks and video
session traffic is intermixed with non-video cross-traffic.  In a real
deployment, the models must identify the video sessions accurately, especially
given that errors in identifying applications can propagate to the quality of the
prediction models.  Second, the prediction models should apply to a range of
services, which existing models tend not to do.  A model that can predict
quality across multiple services is hard because both video streaming
algorithms and content characteristics can vary significantly across video
services (\eg, buffer-based~\cite{huang2014buffer} versus
throughput-based~\cite{stockhammer2011dynamic} rate adaption algorithm,
fixed-size~\cite{huang2014buffer} versus variable-size video
segments~\cite{mondal2017candid}).

This work takes a step towards making video inference models practical,
tackling the challenges that arise when the models must operate on
real network traffic traces and across a broad range of services.  As a proof
of concept that a general model that applies across a range of services in a
real deployment can be designed and implemented, we studied four major
streaming services---Netflix, YouTube, Amazon, and Twitch---across a 16-month
period, in 66 home networks in the United States and France, comprising a
total of 216,173 video sessions. To our knowledge, this deployment study is
the largest public study of its kind for video quality inference.

We find that models that are trained across all four services ({\em composite}
models) perform almost as well as the service-specific models that were
designed in previous work, provided that the training data contains traffic
from each of the services for which quality is being predicted. On the other
hand, we fall short of developing a truly general model that can predict video
quality for services that are not in the training set. An important challenge
for future work will be to devise such a model. Towards the goal of developing
such a model, we will release our training data to the community, which
contains over 13,000 video sessions labeled with ground truth video quality,
as a benchmark for video quality inference, so that others can compare against our work and
build on it.

Beyond the practical models themselves, the deployment study that we performed
as part of this work has important broader implications for both ISPs
and consumers at large.  For example, our deployment study reveals that the
speeds that consumers purchase from their ISPs have
considerably diminishing returns with respect to video quality. Specifically,
Internet speeds higher than about 100~Mbps of downstream throughput offer only
negligible improvements to video quality metrics such as startup delay and
resolution. This new result raises important questions for operators and
consumers. Operators may focus on other aspects of their networks to optimize
video delivery; at the same time, consumers can be more informed about what
downstream throughput they actually need from their ISPs to achieve acceptable application quality.

The rest of the paper proceeds as follows. Section~\ref{sec:background}
provides background on streaming video quality inference and details the state
of the art in this area; in this section, we also demonstrate that previous
models do not generalize across a range of services.
Section~\ref{sec:metrics} details the video quality metrics we aim to predict,
and the different classes of features that we use as input to the prediction
models. Section~\ref{sec:model-def} explains how we selected the appropriate
prediction model for each video quality metric and how we trained the
models. Section~\ref{sec:models} discusses the model validation and
Section~\ref{sec:results} describes the results from the 16-month, 66-home
deployment.  Section~\ref{sec:conclusion} concludes with a summary and
discussion of open problems and possible future directions.

\section{Background and Related Work: \\ Streaming Video Quality
Inference}\label{sec:background}

In this section, we first provide background on DASH video streaming. Then, we
discuss the problem of streaming video quality inference that ISPs face, and
the current state of the art in video quality inference.

Internet video streaming services typically use Dynamic Adaptive Streaming
over HTTP (DASH)~\cite{sodagar2011mpeg} to deliver a video stream. DASH
divides each video into time intervals known as {\em segments} or chunks (of possibly
equal duration), which are then encoded at multiple bitrates and resolutions.
These segments are typically stored on multiple servers (\eg, a content
delivery network) to allow a video client
to download segments from a nearby server.
At the beginning of a video session, a client downloads a DASH Media
Presentation Description (MPD) file from the server. The MPD file contains all
of the information the client needs to retrieve the video (\ie, the audio and
video segment file information). The client sequentially issues HTTP requests
to retrieve segments at a particular bitrate. An {\em application-layer
Adaptive Bitrate (ABR) algorithm} determines the quality of the next segment
that the client should request. ABR algorithms are proprietary, but most video
services rely on recently experienced bandwidth~\cite{stockhammer2011dynamic},
current buffer size~\cite{huang2014buffer}, or a hybrid of the
two~\cite{yin2015control} to guide the selection. The downloaded video
segments are stored in a client-side application buffer. The video buffer is
meant to ensure continuous playback during a video session. Once the size of
the buffer exceeds a predefined threshold, the video starts playing.

ISPs must infer video quality based on the observation of the packets traversing the
network. 
The following video quality metrics affect a user's quality of
experience~\cite{dobrian2011understanding,
balachandran2012quest,balachandran2013developing,ahmed2017suffering,akamaidriving,ITUDoc,krishnan2013video}:
\begin{itemize}
    \itemsep=-1pt
    \item {\em Startup delay} is the time elapsed from the moment the player initiates a
connection to a video server to the time it starts rendering video frames.
\item The {\em resolution} of a video is the number of pixels in each dimension of video frame. 
\item The {\em bitrate} of a segment measures the number of bits transmitted
    over time. 
\item {\em Resolution switching} has a negative effect on
quality of experience~\cite{balachandran2012quest,hossfeld2014assessing}; both
the amplitude and frequency of switches affect quality of
experience~\cite{hossfeld2014assessing}.  
\item {\em Rebuffering} also increases
rate of abandonment and reduces the likelihood that a user will return to the
service~\cite{dobrian2011understanding,krishnan2013video}.
\end{itemize}
\noindent
Resolution and bitrate can vary over the
course of the video stream.  Players may switch the bitrate to adapt to
changes in network conditions with the goal to select the best possible
bitrate for any given condition. 
 
In the past, ISPs have relied on deep-packet inspection to identify video
sessions within network traffic and subsequently infer video
quality~\cite{sandvineDPI,mangla2017mimic,trevisan2016impact}.  Yet, most
video traffic is now encrypted, which makes it more challenging to infer video
traffic~\cite{gsma2015,dyer2015}.

Previous work has attempted to infer video quality from encrypted traffic
streams, yet has typically done so (1)~in controlled settings, where the video
traffic is the only traffic in the experiment; (2)~for single services, as
opposed to a range of services. It turns out that these existing methods do
not apply well to the general case, or in deployment, for two reasons. First,
in real deployments, video streaming traffic is mixed with other
application traffic, and thus simply {\em identifying the video sessions}
becomes a challenging problem in its own right. Any errors in identifying the
sessions will propagate and create further errors when inferring any quality
of experience features. Second, different video streaming services behave in
different ways---video on demand services incur a higher startup delay for
defense against subsequent rebuffering, for example. As such, a model that is
trained for a specific service will not automatically perform accurate
predictions for subsequent services.

Previous work has developed models that operate in controlled settings for
specific services that often do not apply in deployment scenarios or for a
wide range of services.  Mazhar and Shafiq~\cite{infocomPaper},
BUFFEST~\cite{krishnamoorthi2017buffest} and Requet~\cite{gutterman2019requet}
infer video quality over short time intervals, which is closest to our 
deployment scenario. The method from Mazhar and
Shafiq~\cite{infocomPaper}  is the most promising for our goals because they infer 
three video quality metrics: rebuffering events, video quality, and startup delay. 
In addition, their models are based on
decision trees, so we can retrain them for the four services we study using
our labeled data for each service, without the need to reverse engineering
each individual service. 
BUFFEST~\cite{krishnamoorthi2017buffest} focused only on rebuffering detection
for YouTube, but introduces a method for identifying video segments that we
build on in this work.  We aim to infer other quality metrics that are more
indicative of user experience, since most services experience problems with
resolution and startup delay, as opposed to rebuffering. 

\begin{figure}[t]
\begin{center}
\includegraphics[width=\linewidth]{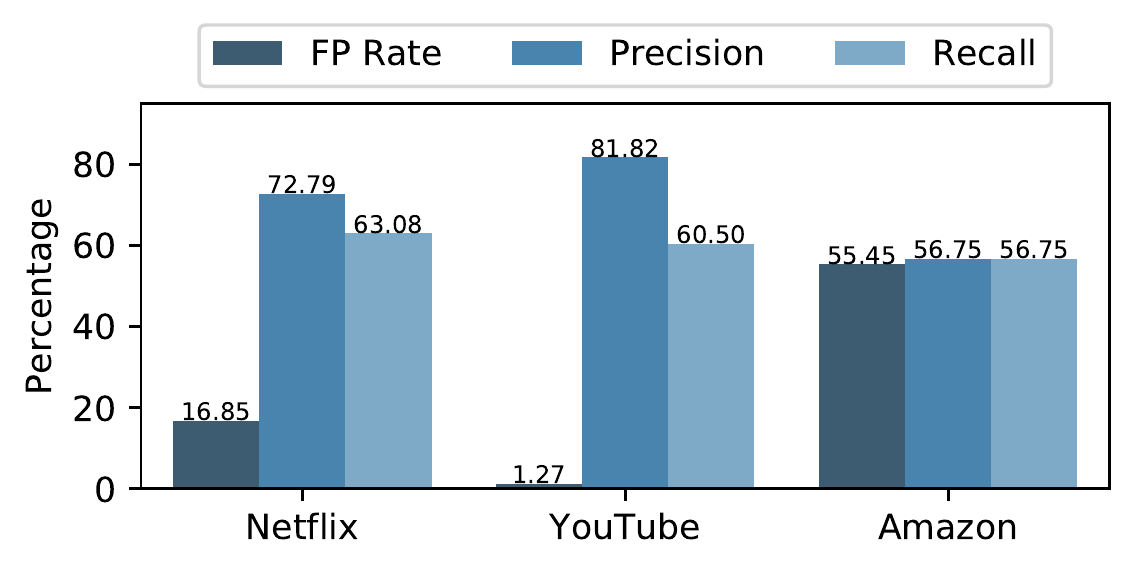}
\caption{Using a per-service model to detect high (bigger than 5 seconds) startup delay
using method from Mazhar and Shafiq~\cite{infocomPaper}.}
\label{fig:adaboost1}
\end{center}
\end{figure}

The shortcomings of these previous approaches are apparent when we attempt to
apply them to a wider range of video services.
When we apply the models from Mazhar and Shafiq to the labeled
datasets we collected with the four video services (described in Section~\ref{sec:trainingdata})
the accuracy was low.  
The models from  Mazhar and Shafiq infer quality metrics at a coarse granularity: 
whether there was rebuffering, whether video quality is high or low, and 
whether the video session has started or not. Here, we show results for the startup 
delay---whether the video started in more than five seconds after the video page has 
been requested. We use Scikit-learn's~\cite{scikit-learn} AdaBoost implementation with one hundred
default weak learners as in the original paper. We train one model per service and evaluate
how well we can detect long startup delays.
Figure~\ref{fig:adaboost1} shows the false positive rate (FPR), precision, and
recall when testing the model for each video service. We omit results for Twitch as 
99\% of Twitch sessions start playing after five seconds, so predicting a startup delay that exceeds five
seconds was not a meaningful exercise---it almost always happens.

Mazhar and Shafiq's method was  ineffective at detecting high startup
delay for the other services. False positive rates are around 20\% for Netflix
and even higher for Amazon. Perhaps a threshold different than five seconds
would be more appropriate for Netflix and Amazon. Instead of focusing on fine
tuning the parameters of their model with the exact same set of features and
inference goals from the original paper, we apply Mazhar and Shafiq's general
approach but revisit the design space. In particular, we define inference
goals that are more fine-grained (\eg, inference of the exact startup delay
instead of whether the delay is above a certain threshold) and we leverage
BUFFEST's method to identify video segments from encrypted traffic to consider
a larger set of input features. 

In addition to problems with accuracy, many previous models have problems with
{\em granularity}. Specifically, many methods infer video quality metrics of
{\em entire video sessions} as opposed to continuous estimates of video sessions
over shorter time intervals, as we do in this work. Continuous estimates of
video quality are often more useful for troubleshooting, or other
optimizations that could be made in real-time.
Dimopoulos \ea~\cite{dimopoulos2016measuring} relied on a web proxy in the
network to inspect traffic and infer quality of YouTube sessions; it is
apparent that they may have performed man-in-the-middle attacks on the
encrypted traffic to infer quality.
eMIMIC~\cite{manglaemimic} relies on the method in
BUFFEST~\cite{krishnamoorthi2017buffest} to identify video segments and build
parametric models of video session properties that translate into quality
metrics.  This model assumes that video segments have a fixed length, which
means that it cannot apply to streaming services with variable-length segments
such as YouTube. 
Requet~\cite{gutterman2019requet} extended BUFFEST's video segment detection
method to infer potential buffer anomalies and resolution at each segment
download for YouTube.  These models rely on buffer-state estimation and are
difficult to apply to other video services, because each video service (and
client) has unique buffering behavior, as the service may prioritize different
features such as higher resolution or short startup delays.

\section{Metrics and Features}\label{sec:metrics}

In this section, we define the problem of video quality inference from
encrypted network traffic. We first discuss the set of video quality metrics
that a model should predict, as well as the granularity of those predictions.
Then, we discuss the space of possible features that the model could use.

\begin{table*}[t]
\centering
\small
\begin{tabular}{|p{5.6cm}|p{5.1cm}|p{5.8cm}|}
\hline
\textbf{Network Layer} & \textbf{Transport Layer} & \textbf{Application Layer}\\% & \textbf{Ground Truth}\\
\hline

throughput up/down (total, video, non-video)\hfill & \# flags up/down (ack/syn/rst/push/urgent)\hfill & segment sizes (all previous, last-10, cumulative)\hfill \\ %& startup delay \\

throughput down difference\hfill & receive window size up/down\hfill & segment requests inter arrivals\hfill \\ % & resolution\\

packet count up/down\hfill  & idle time up/down\hfill & segment completions inter arrivals\hfill \\ % & \# rebufferings\\

byte count up/down\hfill & goodput up/down\hfill & \# of pending request\hfill \\ %& \# quality switches\\

packet inter arrivals up/down\hfill  & bytes per packet up/down\hfill & \# of downloaded segments\hfill \\ %& \\

\# of parallel flows\hfill & round trip time\hfill & \# of requested segments\hfill \\% & \\

 & bytes in flight up/down\hfill & \\%& \\

 & \# retransmissions up/down\hfill & \\%& \\

 & \# packets out of order up/down\hfill & \\

\hline
\end{tabular}
% }
\caption{Summary of the extracted features from traffic.} %Amount of state required: Simple state~(\Circle), Simple application-layer state~(\LEFTcircle), Complex state~(\CIRCLE).}
\label{tab:features}
% \vspace{-3mm}
\end{table*}

\subsection{Target Quality Metrics}

We focus on building models to infer startup delay and resolution.
Prior work has focused on bitrate as a way to approximate the video resolution, but the
relationship between bitrate and resolution is complex because the bitrate
also depends on the encoding and the content type~\cite{aaron2015per}.
Resolution switches can be inferred later from the resolution per time slot.
We omit the models for rebuffering as our dataset only contains rebuferring
for YouTube. We present rebuffering models for YouTube at a technical
report~\cite{bronzino2019lightweight}.

\paragraph{Startup delay.} Prior work presents different models of startup delay.
eMIMIC~\cite{manglaemimic} defines
startup delay as the delay to download the first segment. This definition
can be misleading, however, because most services download multiple segments
before starting playback.
For example, we observe in our dataset that Netflix downloads, on
average, four video segments and one audio segment before playback begins. Alternatively,
Mazhar and Shafiq~\cite{infocomPaper} rely on machine learning to predict
startup delay as a binary prediction~\cite{infocomPaper}.
The threshold to classify a startup delay as short or long, however, varies depending on the
nature of the service, (\eg, roughly three seconds for short videos on a service
like YouTube and five seconds for movies on a service such as Netflix).
Therefore, we instead use a regression model
to achieve finer inference granularity and generalize across video
services.

\paragraph{Resolution.} Resolution often varies through
the course of a video session; therefore, the typical inference target is the
average resolution. Previous work has predominantly used
classification approaches~\cite{dimopoulos2016measuring,infocomPaper}: either
binary (good or bad)~\cite{infocomPaper} or with three classes (high
definition, standard definition, and low
definition)~\cite{dimopoulos2016measuring}. Such a representation can be
misleading, as different clients may have different hardware configurations
(\eg, a smartphone with a 480p screen compared to a 4K smart television).
Therefore, our models infer resolution  as a multi-class classifier,  which is
easier to analyze. Each class of the classifier corresponds to one of the
following resolution values: 240p, 360p, 480p, 720p, and 1080p.

One can compute the average resolution of a video  session~\cite{dimopoulos2016measuring, manglaemimic}
or of a time slot~\cite{infocomPaper}. We opt to infer resolution over time as this fine-grained inference works
for online monitoring plus it can later
be used to recover per-session statistics as well as to infer the frequency and
amplitude of resolution switches. We consider bins of five and ten seconds or at the end of each downloaded video
segment.  To determine the bin size, we study the number of resolution switches
per time bin using our ground truth. The overwhelming majority of
time bins have no bitrate switches. For example, at
five seconds, more than 93\% of all bins for Netflix have no quality switch. As
the bin size increases, more switches occur per bin and
resolution inference occurs on a less precise timescale. At the same time, if
the time bin is too short,
it may only capture the partial download of a video segment, in particular
when the network conditions are poor. We select ten-second bins, which
represents a good tradeoff between the precision of inferring resolution and
the likelihood that each time bin contains complete video segment
downloads.

\subsection{Features}\label{sec:input}
For each video session, we
compute a set of features from the captured traffic at
various levels of the network stack, as summarized in Table~\ref{tab:features}.
We consider a super-set of the features used in prior
models~\cite{dimopoulos2016measuring, infocomPaper, manglaemimic} to
evaluate the sub-set of input features that provides the best inference accuracy.

\paragraph{Network Layer.} We define network-layer features as metrics that
solely rely on information available from observation of a network
flow (identified by the IP/port four-tuple).

Flows to video services fall into three categories: flows that carry video
traffic, flows that belong to the service but transport other type of
information (\eg, the structure of the web page), and all remaining traffic
flows traversing to-and-from the end-host network.  For each network flow
corresponding to video traffic, we compute the upstream and downstream average
throughput as well as average packet and byte counts per second. We also
compute the difference of the average downstream throughput of video traffic
between consecutive time slots. This metric captures temporal variations in
video content retrieval rate. Finally, we compute the upstream and downstream
average throughput for service flows not carrying video and for the total
traffic.

\paragraph{Transport Layer.} Transport-layer features include
information such as end-to-end latency and packet retransmissions. These metrics
reveal possible network problems, such as presence of a lossy link in the path or
a link causing high round-trip latencies between the client and the server. Unfortunately,
transport metrics suffer two shortcomings. First, due to encryption, some metrics
are only extractable from TCP flows and not flows that use QUIC (which is increasingly
prevalent for video delivery). Second, many transport-layer features require maintaining
long-running per-flow state, which is prohibitive at scale. For the metrics in Table~\ref{tab:features},
we compute summary statistics such as mean, median,
maximum, minimum, standard deviation, kurtosis, and skewness.

\paragraph{Application Layer.} Application-layer metrics include any feature
related to the application data, which often provide the greatest insight into
the performance of a video session. Encryption, however, makes it impossible
to directly extract any application-level information from traffic using deep
packet inspection. Fortunately, we can still derive some application-level
information from encrypted traffic. For example,
BUFFEST~\cite{krishnamoorthi2017buffest} showed how to identify individual
video segments from the video traffic, using the times of upstream requests
(\ie, packets with non-zero payload) to break down the stream of downstream
packets into video segments. Our experiments found that this method works well
for both TCP and QUIC video traffic. In the case of QUIC, signaling packets
have non-zero payload size, so we use a threshold for QUIC UDP payload size to
distinguish upstream and downstream packets. Based on observations obtained from a dataset
of YouTube QUIC sessions traffic traces, we set the threshold to 150 Bytes.

We use sequences of inferred video segment downloads to build up the feature
set for the application layer.  For each
one of the features  in Table~\ref{tab:features} we compute the following statistics: minimum, mean,
maximum, standard deviation and 50th, 75th, 85th, and 90th percentile.

\section{Model Selection and Training}\label{sec:model-def}

We describe how we gathered the inputs for the prediction model, how we
selected the prediction model for startup delay and resolution, and the process for training
the final regressor and classifier.

\subsection{Inputs and Labeled Data}\label{sec:trainingdata}

\paragraph{Gathering input features.} We train models considering different
sets of input features: network-layer features (\emph{Net}), transport-layer
features (\emph{Tran}), application-layer features (\emph{App}), as well as a
combination of features from different layers: Net+Tran, Net+App and all
layers combined (\emph{All}). For each target quality metric, we train 32 models in total: (1)~varying
across these six features sets and (2)~using six different datasets,
splitting the dataset with sessions from each of the four video services---Netflix,
YouTube, Amazon, and Twitch---plus two combined datasets, one with
sessions from all services (which
we call \textit{composite}) and one with sessions from three out of four services (which
we call \textit{excluded}). For models that rely on transport-layer
features, we omit YouTube sessions over UDP as we cannot compute all features.
For each target quality metric, we evaluate models using 10-fold
cross-validation.  We do not present the results for models based only on
transport-layer or application-layer features, because the additional cost of
collecting lightweight network-layer features is minimal, and these models are
less accurate, in any case.

\paragraph{Labeling: Chrome Extension.}
To label these traffic traces with the appropriate video quality metrics, we
developed a Chrome extension that monitors application-level information for
the four services. This extension, which supports any HTML 5-based video,
allowed us to assign video quality metrics to each stream as seen by the
client.\footnote{We will release the extension upon publication.}

The extension collects browsing history by parsing events available from the
Chrome WebRequest APIs~\cite{webRequest}. This API exposes all necessary
information to identify the start and end of video sessions, as well as the
HTTPS requests and responses for video segments. To collect video quality
metrics, we first used the Chrome browser API to inspect the URL of every page
and identify pages reproducing video for each of the video services of
interest.  After the extension identifies the page, collection is tailored for
each service:

\begin{itemize}
    \itemsep=-1pt
    \item \textit{Netflix: Parsing overlay text.} Netflix reports video quality
statistics as an overlay text on the video if the user provides a specific
keystroke combination. We injected this keystroke combination but render the
text invisible, which allows us to parse the reported statistics without
impacting the playback experience. This information is updated once per
second, so we adjusted our collection period accordingly. Netflix reports a
variety of statistics. We focused on the player and buffer state information;
including whether the player is playing or not, buffer levels (\ie, length of
video present in the buffer), and the buffering resolution.
\item \textit{YouTube: iframe API.} We used the YouTube iframe API~\cite{iframe}
to periodically extract player status information, including current video
resolution, available playback buffer (in seconds) and current playing
position. Additionally, we collect events reported by the \textit{<video>
HTML5} tag, which exposes the times that the player starts or stops the video
playback due to both user interaction (\eg, pressing pause) or due to lack of
available content in the buffer.
\item \textit{Twitch and Amazon: HTML 5 tag parsing.} As the two services
expose no proprietary interface, we generalized the module developed for
YouTube to solely rely on the \textit{<video> HTML5} tag to collect all the
required data. This approach allowed us to collect all the events described
above as well as player status information, including current video
resolution, available playback buffer (in seconds), and current playing
position.
\end{itemize}

\subsection{Training}

\paragraph{Startup delay.} We trained on features computed from the
first ten seconds of each video session. We experimented with different
regression methods, including: linear, ridge, SVR, decision tree regressor,
and random forest regressor. We evaluate methods based on the average absolute
error and conclude that random forest leads to lowest errors. We select the
hyper parameters of the random forest models on the validation set using the
R2 score to evaluate all combinations with exhaustive grid search.

\paragraph{Resolution.}  We trained a classifier with
five classes: 240p, 360p, 480p, 720p, and 1080p. We evaluated Adaboost (as in
prior work~\cite{infocomPaper}), logistic regression, decision trees, and
random forest. We select random forests because it again gives higher precision and
recall with lower false positive rates.  Similarly to the model for startup
delay, we select hyper parameters using exhaustive grid search and select the
parameters that maximize the F1 score.

\paragraph{Generating and labeling traffic traces.}
We
instrumented 11 machines to generate video traffic and collect packet traces
together with the data from the Chrome extension: six laptops in residences
connected to the home WiFi network (three homes in a large European city with
download speeds of 100~Mbps, 18~Mbps, and 6~Mbps, respectively; one room in a
student residence in the same city; one apartment in a university campus in
the US; and one home from a rural area in the US), four laptops located in our
lab connected via the local WiFi network, and one desktop connected via
Ethernet to our lab network.
We generated video sessions automatically using
\texttt{ChromeDriver}~\cite{ChromeDriver}. We played each session for 8 to
12~minutes depending on the video length and whether there were ads at the
beginning of the session (in contrast to previous work~\cite{infocomPaper}, we
did \textit{not} remove ads from the beginning of sessions in order to recreate the most
realistic setting possible). For longer videos (\eg, Netflix movies), we
varied the playback starting point to avoid always capturing the first portion
of the video. We generate five categories of sessions: Netflix, Amazon, Twitch,
YouTube TCP, and YouTube QUIC. We randomly selected Netflix and Amazon movies
from the suggestions presented by each service in the catalog page. To avoid
bias, we selected movies from different categories including action, comedies,
TV shows, and cartoons. We ultimately selected 25 movies and TV shows used in
rotation from Netflix and 15 from Amazon. Similarly for YouTube,
we select 30 videos from different categories. Twitch automatically starts a
live video feed when opening the service home page. Thus, we simply collect data
from the automatically selected feed. To allow Netflix to stream 1080p content
on Chrome, we installed a dedicated browser extension~\cite{chrome1080}.
Finally, we collected packet traces using \texttt{tcpdump}~\cite{tcpdump} on
the network interface the client uses and compute traffic features presented in
the previous section.

\paragraph{Emulating Network Conditions.}
In the lab environment, we manually varied the network conditions in the
experiments using \texttt{tc}~\cite{tc} to ensure that our training datasets
captured a wide range of network conditions. These conditions can either be
stable for the entire video session or vary at random time intervals. We varied
capacity from 50~kbps to 30~Mbps, and introduced loss rates between 0\% and 1\%
and additional latency between 0~ms and 30~ms. All experiments within homes ran
with no other modifications of network conditions to emulate realistic home
network conditions.

\begin{table}[t]
\small
\begin{tabular}{|l|r|rr|}
\hline
\textbf{Service} & \textbf{Total Runs} &  \textbf{\% Home} & \textbf{\% Lab}\\
\hline
\textbf{Netflix} & 3,861 & 53\% & 47\% \\ %3,498
\textbf{YouTube TCP} & 4,511 & 16\% & 74\% \\ %4,232
\textbf{YouTube QUIC} & 1,310 & 58\% & 42\% \\ %1,101
\textbf{Twitch} & 2,231 & 17\% & 83\% \\
\textbf{Amazon} & 1,852 & 10\% & 90\% \\
% \textbf{Hulu} & 1,086 & 100\% & - \\
\hline
\end{tabular}
\caption{Summary of the labeled dataset.}
\label{table:runs}
\end{table}

\paragraph{Measurement period.}
We collected data from November 20, 2017 to May~3, 2019. We filtered any
session that experienced playing errors during the execution. For example, we
noticed that when encountering particularly challenging video conditions (\eg,
50~kbps download speeds), Netflix's player simply stops and shows an error
instead of stalling and waiting for the connectivity to return to playable
conditions. Hence, we removed all sessions that required more than a high
threshold, \ie, 30 seconds, to start reproducing the video. The resulting
dataset contains 13,765 video sessions from Netflix, Amazon, Twitch, and
YouTube, that we use to train and test our inference models.\footnote{We will release the collected dataset upon publication.}
Table~\ref{table:runs} shows the number of the runs per video service and
under the different network conditions.

\section{Model Validation}\label{sec:models}

We evaluate the accuracy of models relying on different features sets for predicting
startup delay and resolution. Our results show that models that rely on network-
and application-layer features  outperform models that rely on network- and
transport-layer features across all services. This result is in contrast with prior
work~\cite{dimopoulos2016measuring,infocomPaper}, which
provided models that rely on transport-layer features.

We also evaluate whether our models are general. A composite model---where we train
the model with data from multiple services and later predict quality of any
video service---is ideal as it removes the requirement to collect data with
ground truth for a large number of services. Our evaluation of a
\textit{composite model}
trained with data from the four video services shows that it
performs nearly as well as \textit{specific models} that rely only on sessions from a single
service across both quality metrics. This result raises hopes that the composite
model can generalize to a wide variety of video streaming services. When we
train models using a subset of the services and evaluate it against the left out one (\textit{excluded models}),  however, the accuracy of both
startup delay and resolution models degrades significantly, rendering the
models unusable. This result highlights that although our modeling method is general in that it
achieves good accuracy across four video services, the training set used to infer quality metrics should include all services that one aims to do prediction for.

\begin{figure}[t]
\begin{center}
\includegraphics[width=\linewidth]{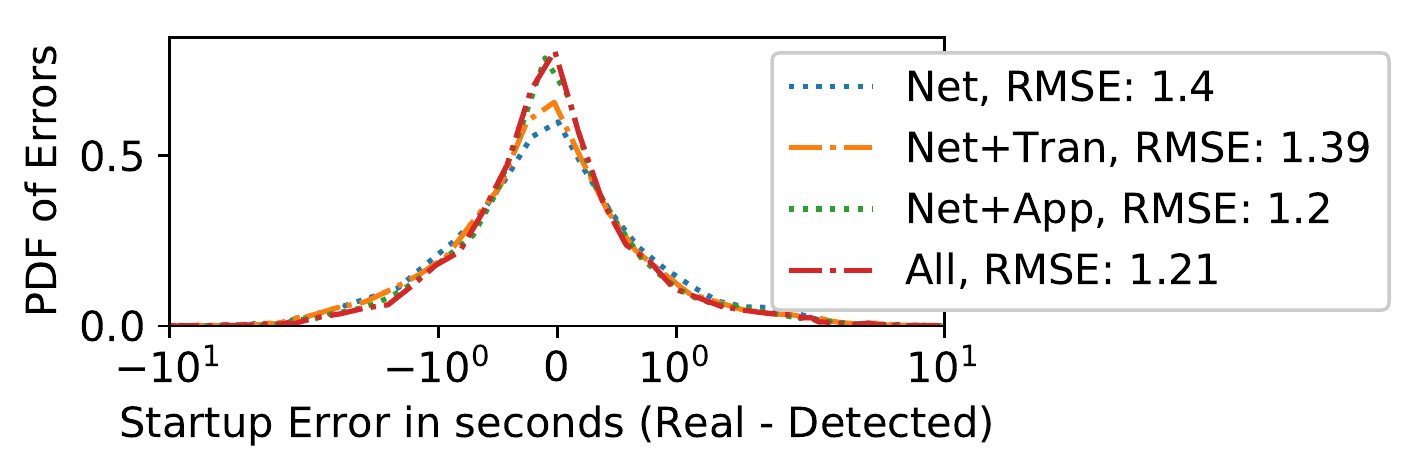}
\caption{Startup delay inference error across different features sets.}
\label{fig:sd_error_bars}
\end{center}
\end{figure}

\subsection{Startup Delay}\label{sec:startup}

To predict startup delay, we train one random forest regressor for each features set and service combination.
We  study the magnitude of the inference errors in our model prediction.
Figure~\ref{fig:sd_error_bars} presents the error of startup delay inference
across the four services for different features sets.
Interestingly, our results  show that the model using a combination
of features from network and application layer yields the highest precision, minimizing
the root mean square error (RMSE) across the four services. These results also show that we
can exclude Net+Tran models, because  Net+App models have consistently smaller errors.
ISPs may ultimately choose a model that uses only network-layer
features, which rely on features that are readily available in many monitoring systems
for a small decrease in inference accuracy.

\begin{figure}[t]
\begin{center}
\includegraphics[width=\linewidth]{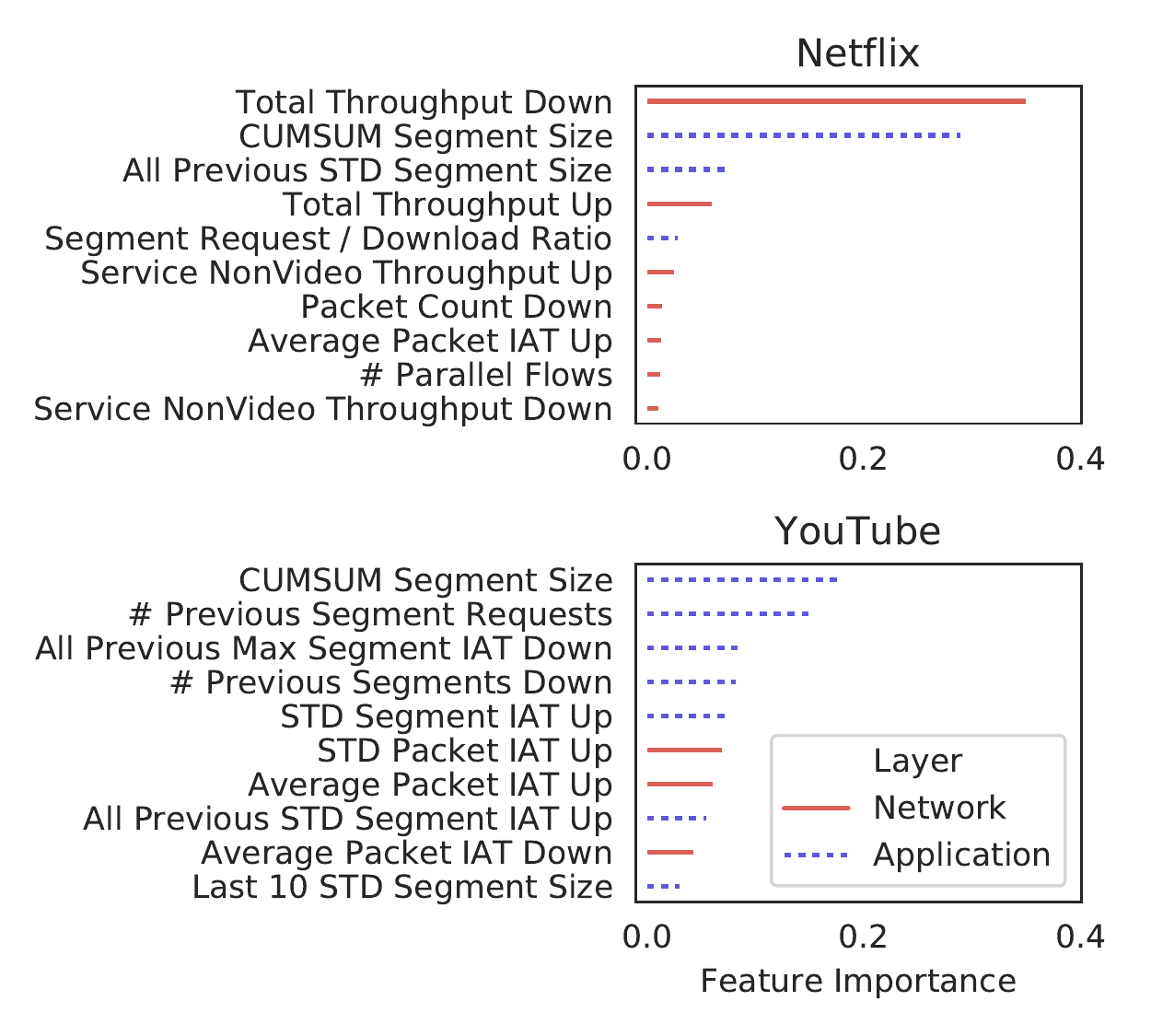}
\caption{Startup delay feature importance for Netflix and YouTube Net+App models.}
\label{fig:feature_sd_important}
\end{center}
\end{figure}

To further understand the effect of different types of features on the models,
we study feature importance (based on the Gini
index~\cite{breiman2017classification}) across the different services. Figure
\ref{fig:feature_sd_important} ranks the ten most important features for both Netflix
and
YouTube for the Net+App model.  Overall, application-layer features dominate
YouTube's ranking with predominantly features that indicate how fast the client
downloads segments, such as the cumulative segment size and their interarrival
times. We observe the same trend for Amazon and Twitch. In contrast, the total
amount of downstream traffic (a network-layer feature) dominates Netflix's
list. Although this result seems to suggest a possible difference across services, this
feature also indicates the number of segments  being
downloaded. In fact, all of these features align with the expected DASH video
streaming behavior: the video startup delay represents the time for filling the
client-side buffer to a certain threshold for the playback to begin. Hence, the faster
the client reaches this threshold (by downloading segments quickly), the lower the
startup delay will be.

\begin{figure*}[t!]
\begin{minipage}{1\linewidth}
\begin{subfigure}[b]{0.24\linewidth}
  \includegraphics[width=\linewidth]{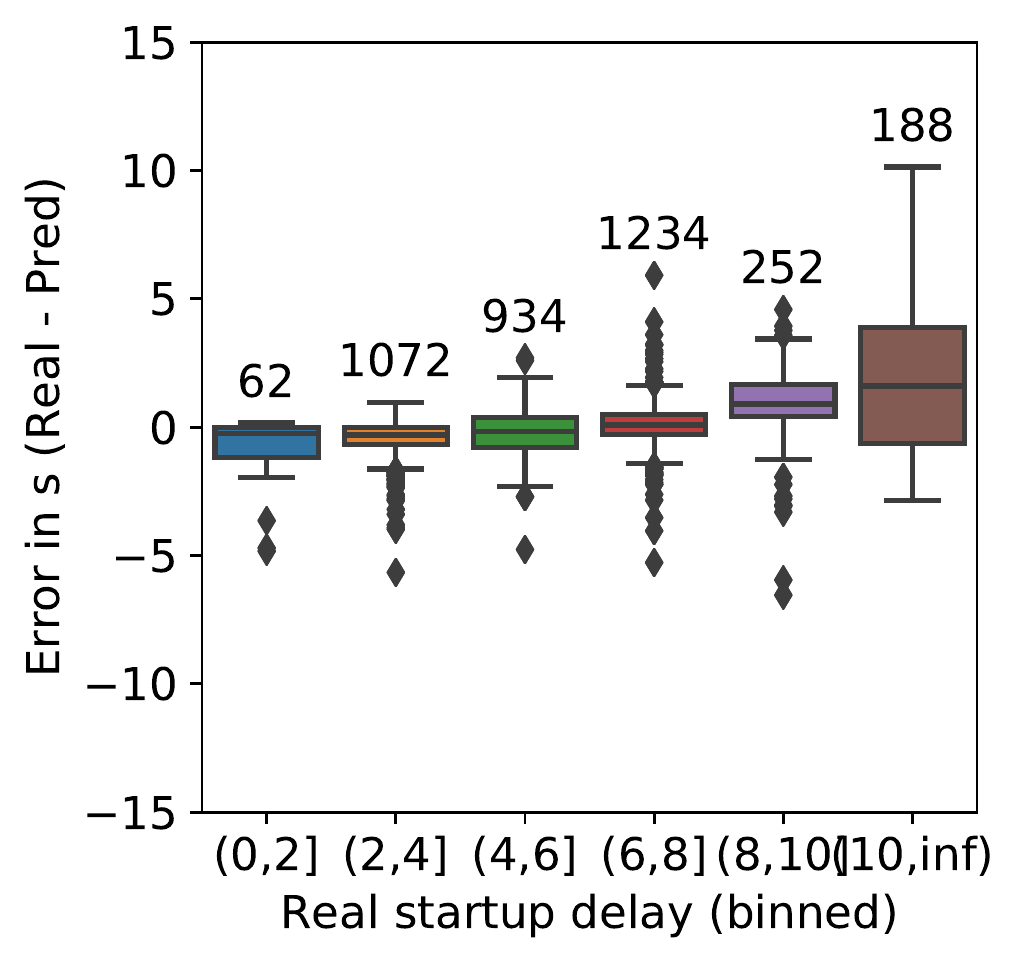}
  \caption{Netflix.}
  \label{fig:sd_netflix}
\end{subfigure}%
\hfill
\begin{subfigure}[b]{0.24\linewidth}
  \includegraphics[width=\linewidth]{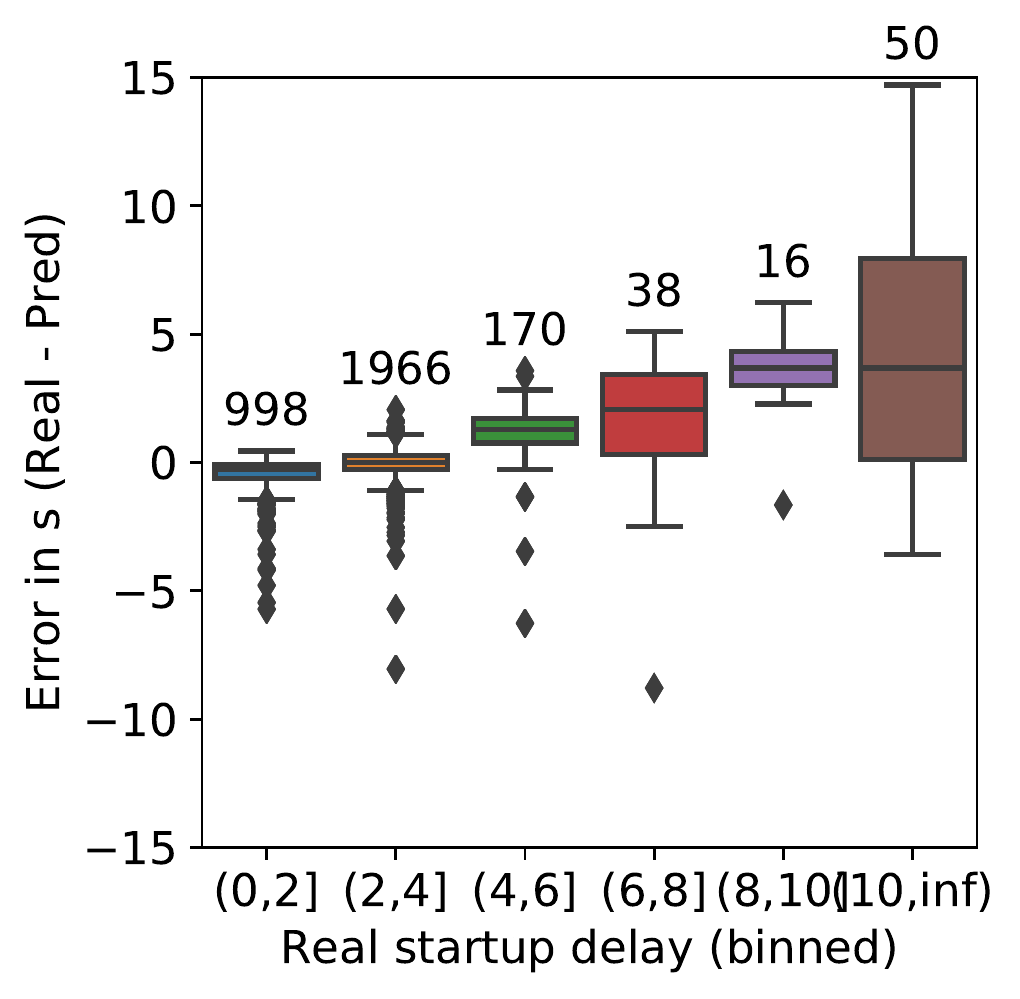}
  \caption{YouTube.}
  \label{fig:sd_youtube}
\end{subfigure}
\hfill
\begin{subfigure}[b]{0.24\linewidth}
  \includegraphics[width=\linewidth]{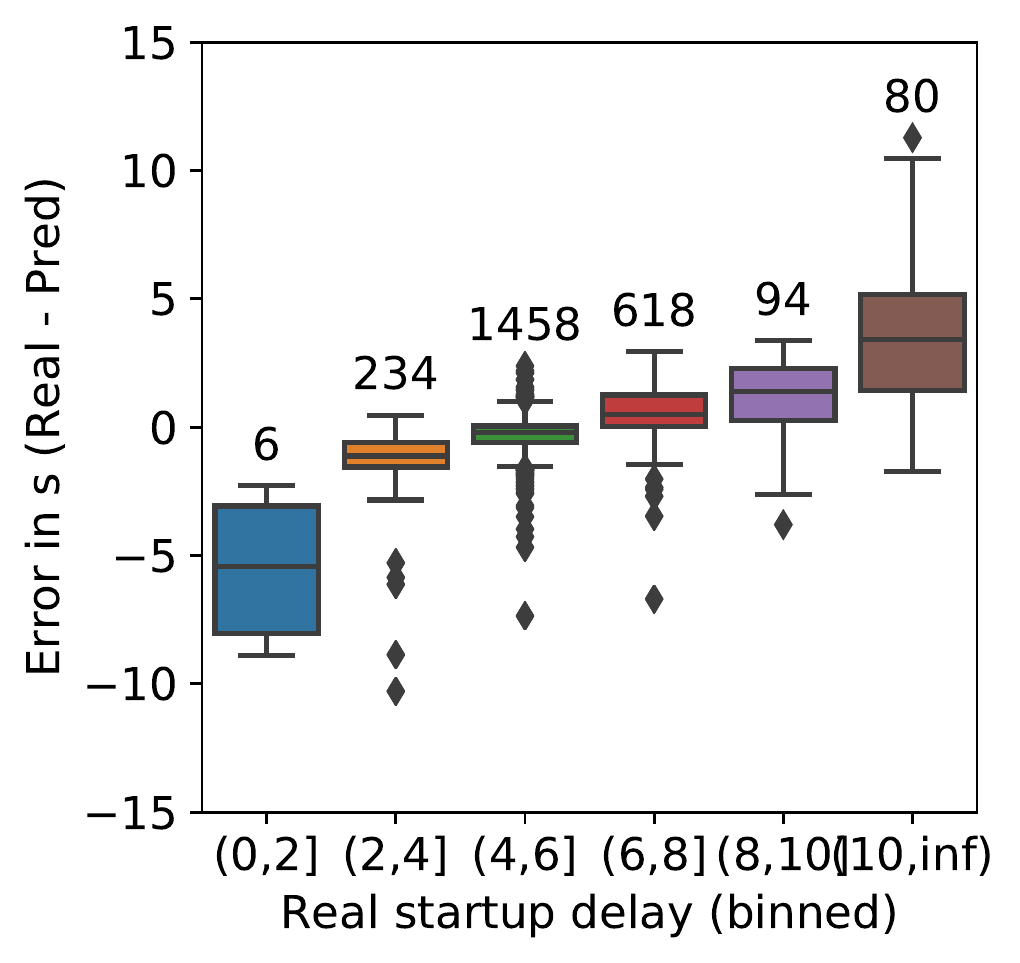}
  \caption{Amazon.}
  \label{fig:sd_amazon}
\end{subfigure}
\hfill
\begin{subfigure}[b]{0.24\linewidth}
  \includegraphics[width=\linewidth]{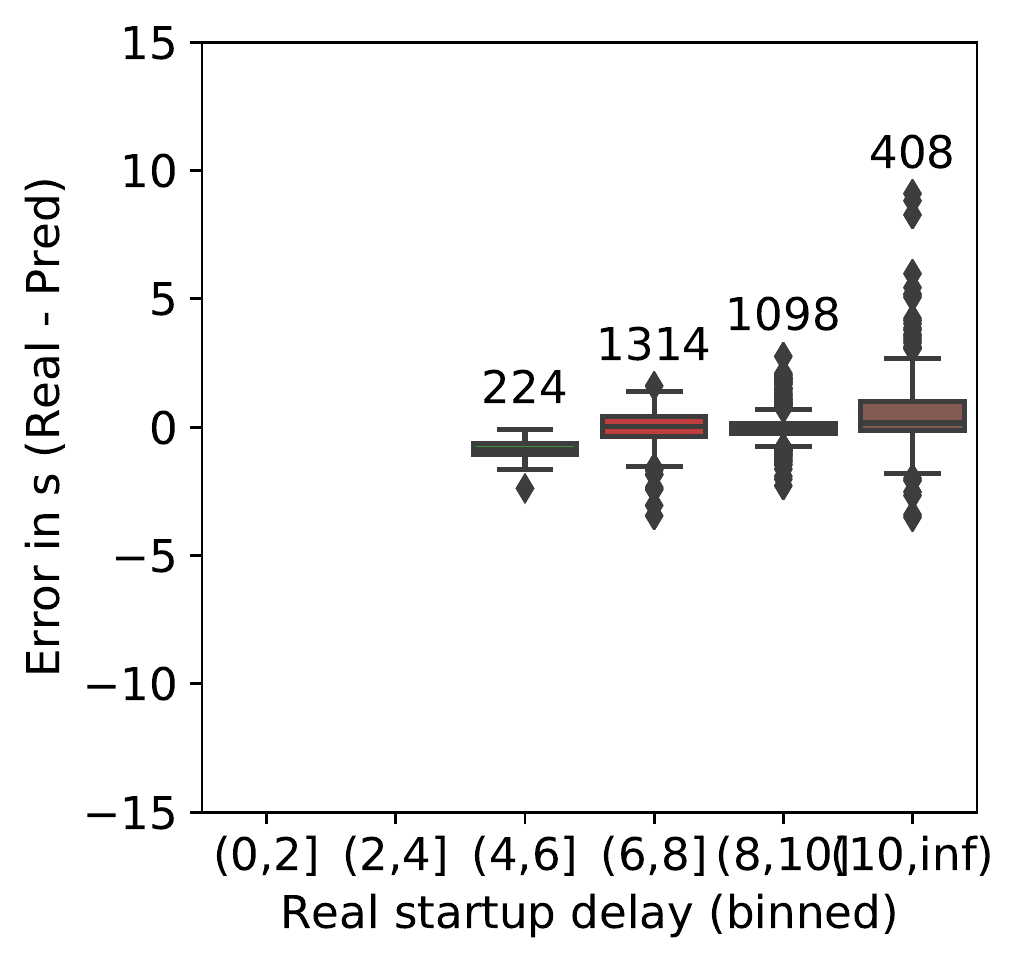}
  \caption{Twitch.}
  \label{fig:sd_twitch}
\end{subfigure}
\end{minipage}
\caption{Startup delay inference error across different services for Net+App specific models.}
\label{fig:service_sd_pdf}
\end{figure*}

Finally, we aim to understand how much the larger errors observed in
Figure~\ref{fig:sd_error_bars} could negatively impact real world applications
of the model for the different services. Figure~\ref{fig:service_sd_pdf} shows for
each service the distribution of the relative errors for inferred startup delays
split into two-second bins (using the Net+App features set). We see that the
errors depend heavily on the number of samples in each bin (displayed
at the top of each box): the larger the
number of samples, the smaller the errors.  This result is expected
as the model better learns the behavior of video services with more data
points, and is also reassuring as the model is more accurate for the cases
that one will more likely encounter in practice. The bins with more
samples have less than one second error most of the time. For example, the
vast majority of startup delays for YouTube occur in the zero to four seconds
bins (91\% of total sample) and errors are within one second for the
majority of these instances. Overall, our models perform particularly well for startup
delays of less than ten seconds with errors mostly within one second (with the
exception of Amazon, which only has six samples in the (0,2]
range of startup delay). Above ten seconds the precision degrades due to
the lower number of samples.

\begin{figure}[t]
\begin{center}
\includegraphics[width=\linewidth]{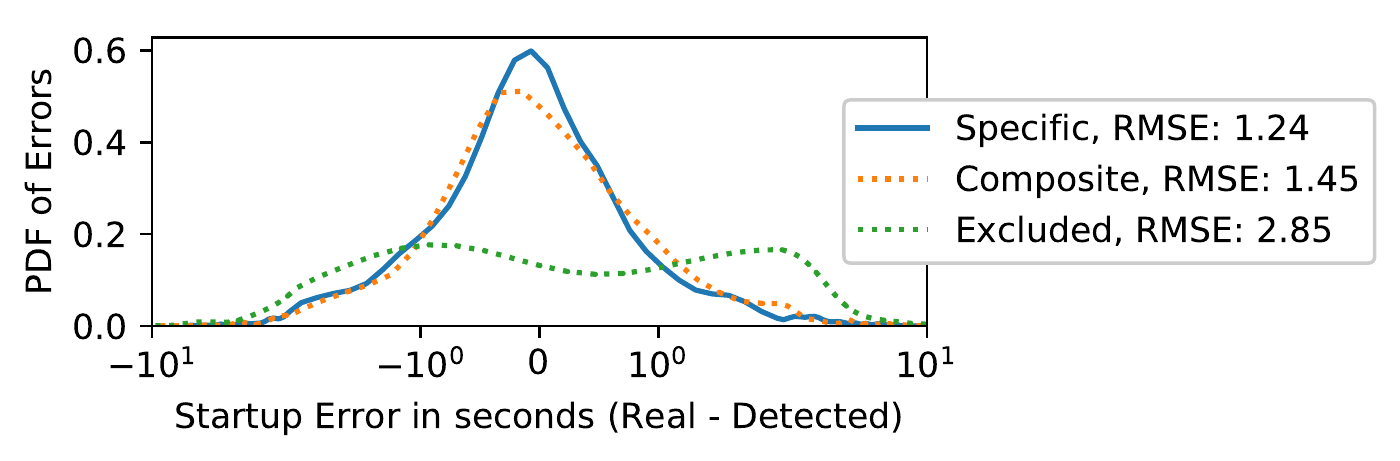}
\caption{Relative error in startup delay inference with same or composite Net+App models.}
\label{fig:startup_bp}
\end{center}
\end{figure}

\paragraph{Composite models.}
We evaluate a composite model for inferring startup delay across multiple services.
Figure~\ref{fig:startup_bp} reports startup delay inference errors for
Netflix for the model trained using solely
Netflix sessions (specific model), the one using data from all four services (composite model), and finally the one using only
the other three services (excluded model). The composite model is helpful in that we can perform training and parameter tuning once
across all the services and deploy a single model online for a small loss in accuracy. We present results for Netflix, but the conclusions were similar
for the other three services. The composite model performs almost as well
as the specific model obtained from tailored training of a regressor for Netflix; the RMSE of the specific
model is 1.24 and that of the composite model is 1.45.  Our analysis of the data shows that errors are
mostly within one second for startup delays  between two and ten
seconds, the biggest discrepancies occur for delays below two and above ten seconds due to the low number
of samples. In summary, these results suggest it is possible to predict startup delay with a composite model across
multiple services.

\begin{figure}[t!]
\begin{minipage}{1\linewidth}
\centering
\begin{subfigure}[b]{0.47\linewidth}
\includegraphics[width=\linewidth]{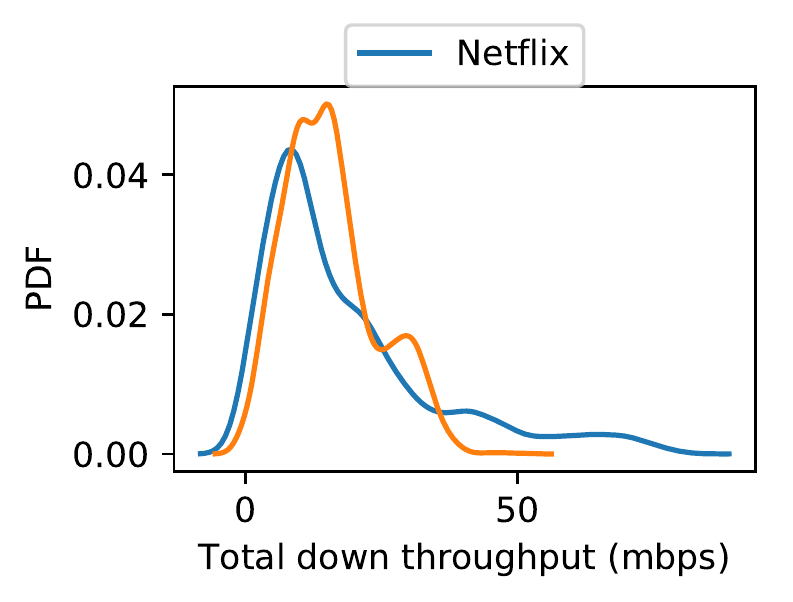}
\caption{Total down throughput.}
\label{fig:cumsum}
\end{subfigure} \hfill
\begin{subfigure}[b]{0.47\linewidth}
\includegraphics[width=\linewidth]{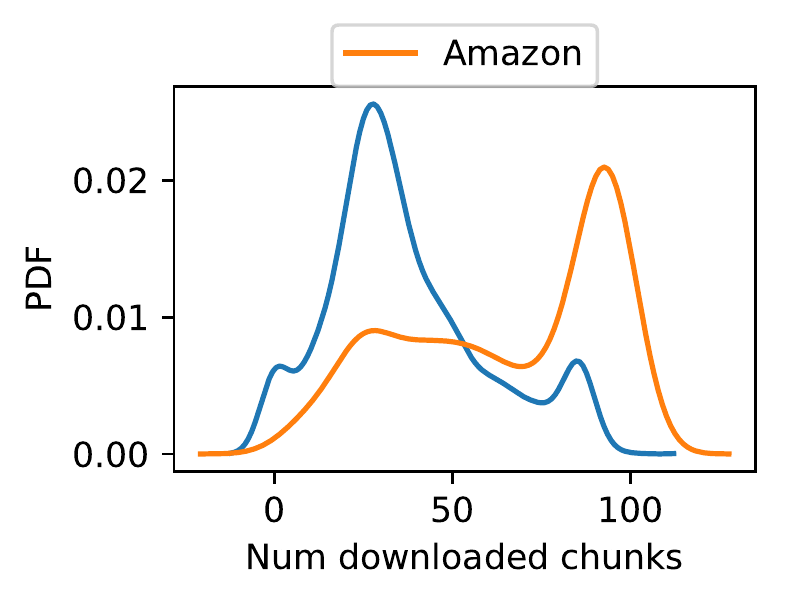}
\caption{Number of downloaded segments.}
\label{fig:std}
\end{subfigure}
\end{minipage}
\caption{Total down throughput is similar across services; number of downaloded segments is not.}
\label{fig:features_comp_startup}
\end{figure}

\paragraph{Applying a composite model to new services.}
Given that all the video streaming services use DASH, our hope was to train the model with data from a few
services and later use it for other services not present in the training set. To test the generality of the model, we evaluate a
model for inferring startup delay, where we train with Amazon, YouTube, and Twitch data and test with Netflix
(excluded model).
Unfortunately, the results in Figure~\ref{fig:startup_bp} show that the excluded model performs very poorly compared to both the
specific and composite models more than doubling the RMSE to 2.91. To better understand this result, we analyze the
differences and similarities of the  values of input features
among the four different services. Although Netflix's behavior is often similar to that of Amazon, the similarities
are concentrated on only a sub-set of the features in our models. For example, Figure~\ref{fig:cumsum}  shows the distribution
of the total downstream throughput in the first ten seconds and Figure~\ref{fig:std} shows the number of downloaded
segments---two of the most important features for Netflix's inference. We observe that the distribution of the downstream
throughput is fairly similar between Netflix and Amazon. On the other hand, the distributions of the number of
downloaded segments present significant differences peaking respectively at 25 and 100 segments. Given these discrepancies, the models
have little basis to predict startup delay
for Netflix when trained with only the other three services. Whether it is possible to build a general model to infer
startup delay across services is an interesting question that deserves further investigation. In the rest of this paper,
we rely on the composite
models based on the Net+App features to infer startup delay across services.

\subsection{Resolution}\label{sec:resolution}

\begin{figure}[t!]
\begin{minipage}{1\linewidth}
\centering
\begin{subfigure}[b]{\linewidth}
\includegraphics[width=\linewidth]{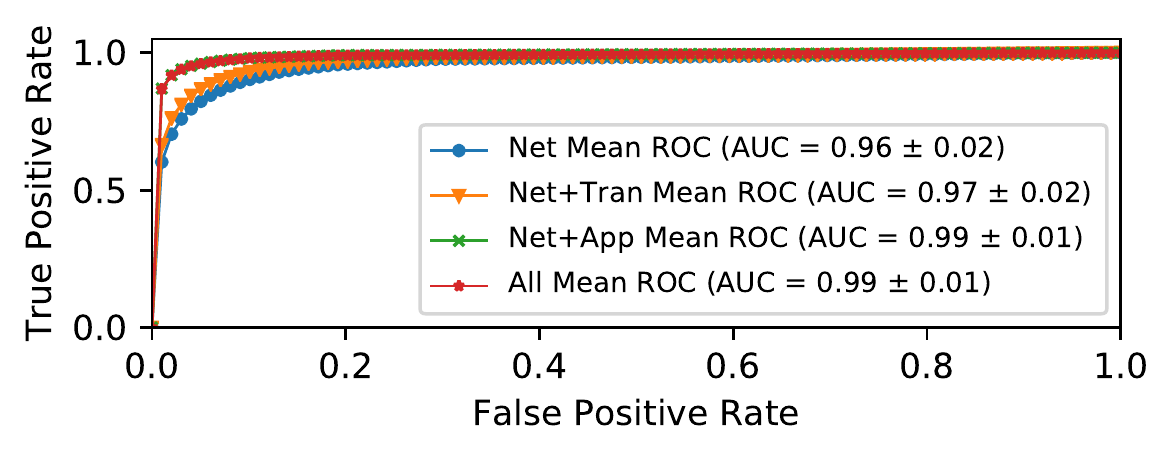}
\caption{ROC.}
\label{fig:feature_res_roc}
\end{subfigure} \hfill
\begin{subfigure}[b]{\linewidth}
\includegraphics[width=\linewidth]{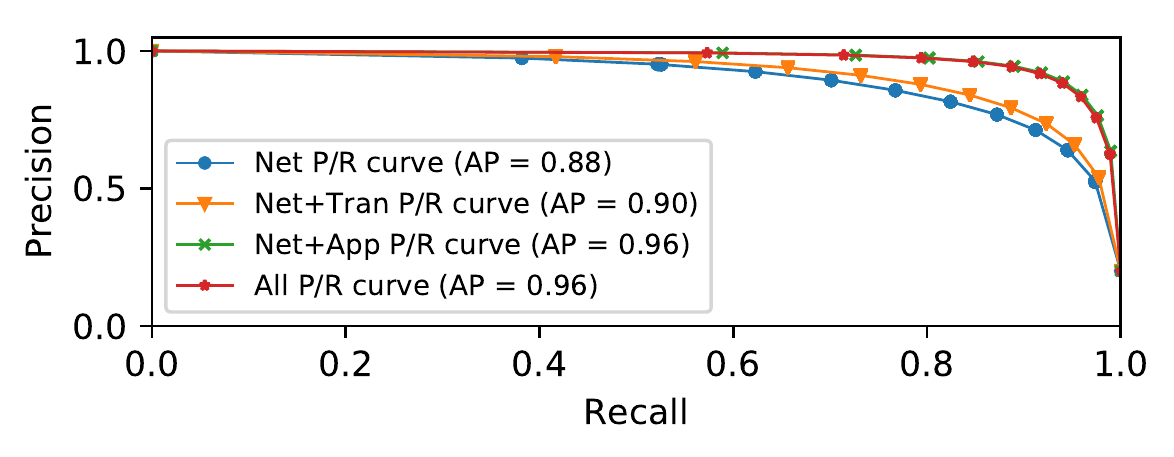}
\caption{Precision-recall.}
\label{fig:feature_res_pr}
\end{subfigure}
\end{minipage}
\caption{Resolution inference using different features sets (For all four video services).}
\label{fig:inference_res}
\end{figure}

\begin{figure}[t]
\begin{center}
\includegraphics[width=\linewidth]{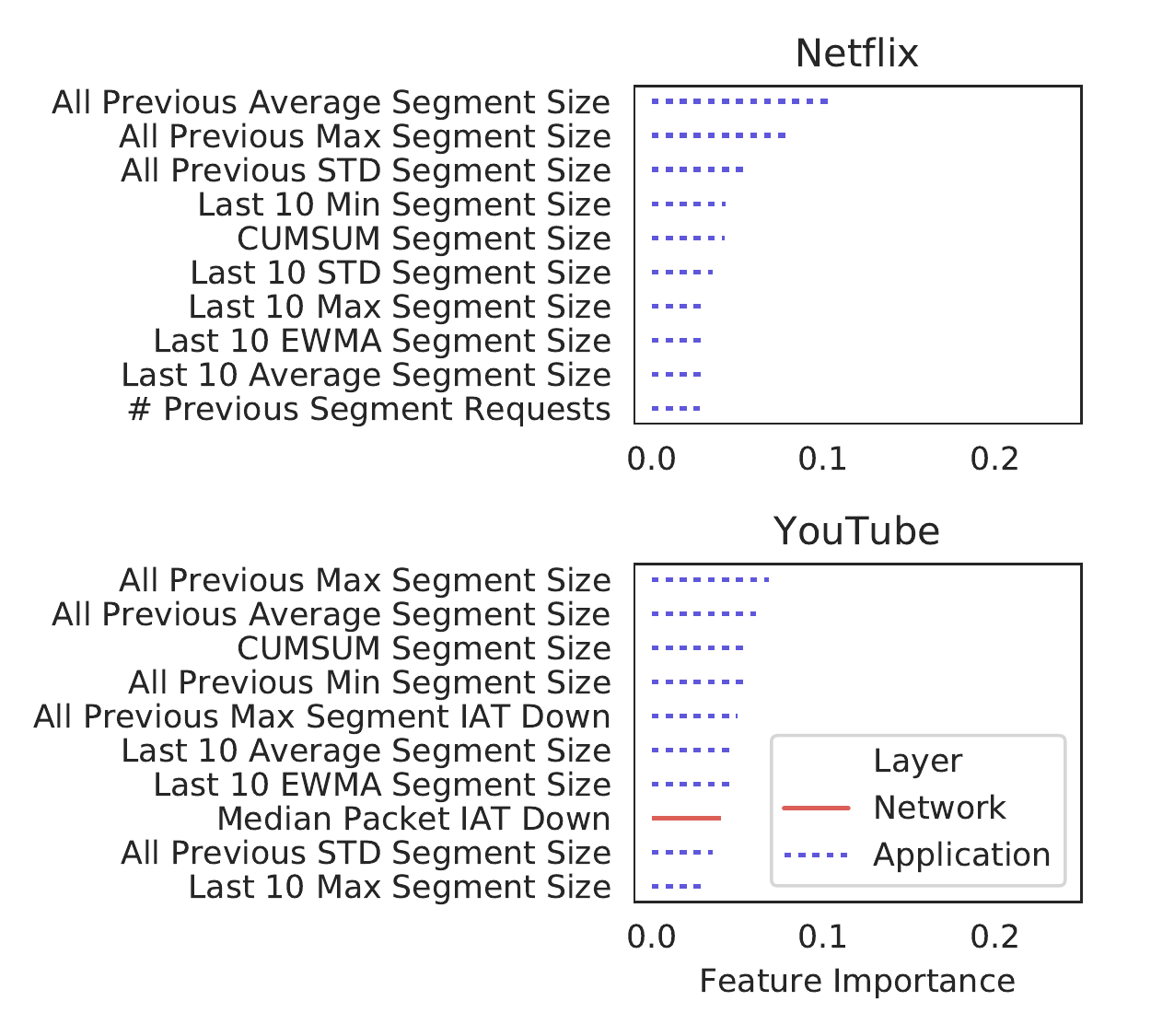}
\caption{Resolution inference feature importance for Netflix and YouTube Net+App models.}
\label{fig:feature_res_important}
\end{center}
\end{figure}

Next, we explore resolution inference. We divide each video session into ten-second
time intervals and conduct the inference on each time bin. We train one random forest
multi-class classifier for each service features set and service combination. Similarly to
the previous section, we present aggregated results for the different layers
across the studied services. We report the receiver operating characteristic
(Figure~\ref{fig:feature_res_roc}) and precision-recall
(Figure~\ref{fig:feature_res_pr}) curves for the models weighted average across the
different resolution labels to illustrate the performance of the classifier for different
values of the detection threshold. The models trained with
application layer features consistently achieve the best performance with both
precision and recall reaching 91\% for a 4\% false positive rate. Any
model not including application features reduces precision by at least 8\%,
while also doubling the false-positive rate.

We next investigate the relative importance of features in the Net+App model.
Figure~\ref{fig:feature_res_important} reports the results for
YouTube and Netflix, which show that most features are related to
segment size. The same applies to the other services. This result confirms
the general intuition: given similar content, higher resolution implies more pixels-per-inch,
thereby requiring more data to be delivered per video segment. Models trained
with network and transport layer features infer resolution by relying on
attributes related to byte count, packet count, bytes per packet,
bytes in flight, and throughput. Without segment-related features, these models
achieve comparatively lower precision and recall.

\begin{figure}[t!]
\begin{minipage}{1\linewidth}
\centering
\begin{subfigure}[b]{\linewidth}
\includegraphics[width=\linewidth]{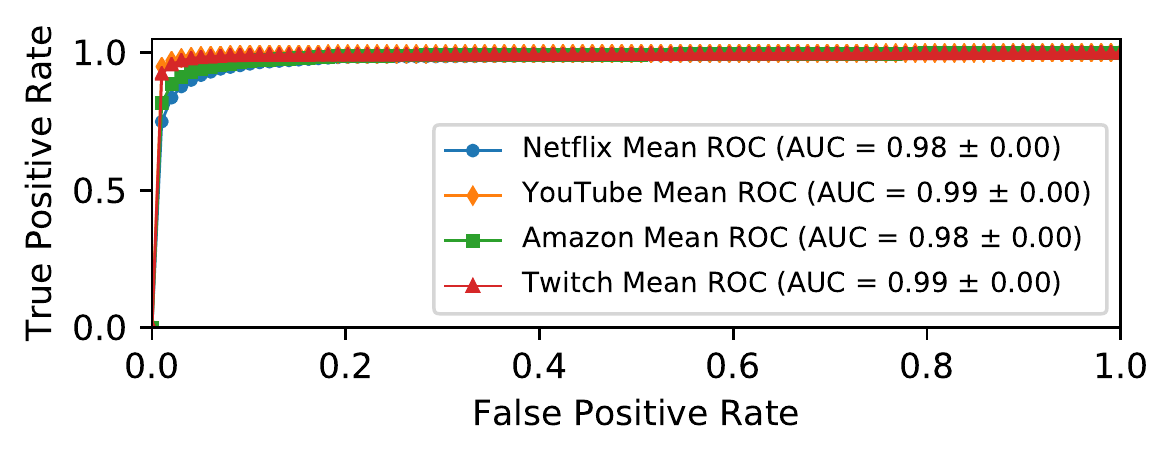}
\caption{ROC.}
\label{fig:service_res_roc}
\end{subfigure} \hfill
\begin{subfigure}[b]{\linewidth}
\includegraphics[width=\linewidth]{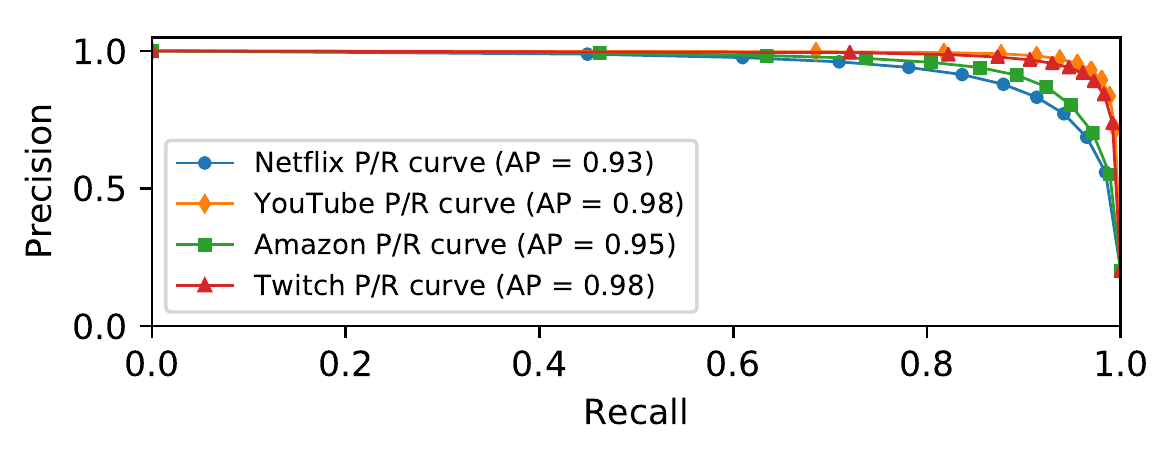}
\caption{Precision-recall.}
\label{fig:service_res_pr}
\end{subfigure}
\end{minipage}
\caption{Resolution inference across services (with Net+App features set).}
\label{fig:feature_res}
\end{figure}

\begin{figure}[t]
\begin{center}
\includegraphics[width=0.9\linewidth]{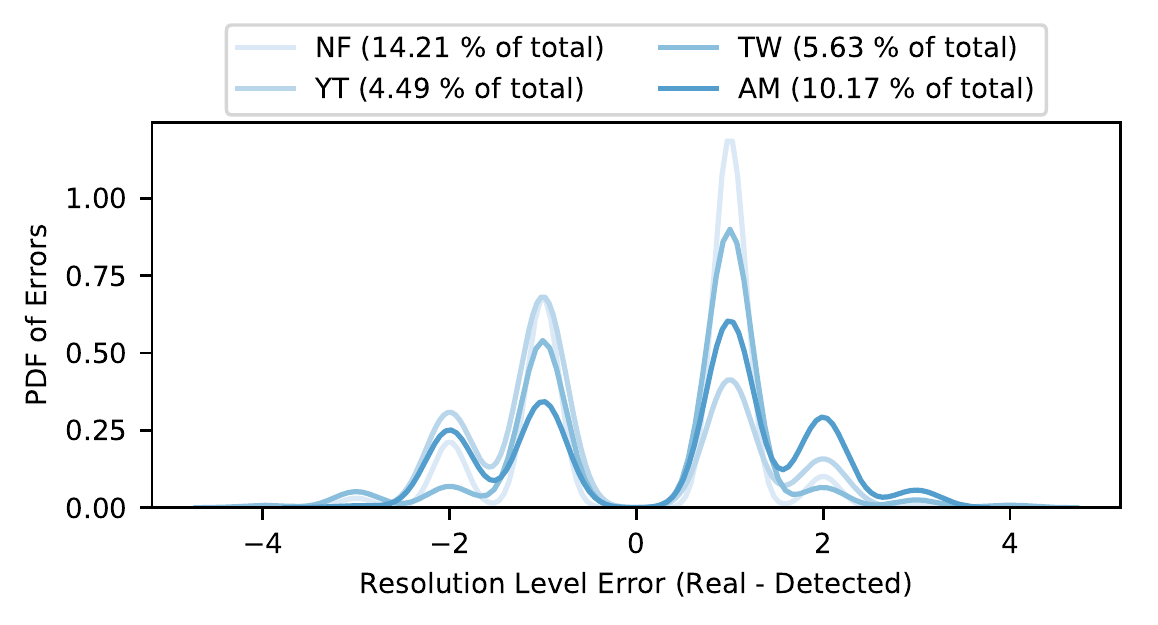}
\caption{Kernel Density Estimation function for the relative error in resolution inference for Net+App models (normalized to levels).}
\label{fig:resolution_rel_error}
\end{center}
\end{figure}

Figure~\ref{fig:feature_res} presents the accuracy for each video service using the
specific model. We see that overall the precision and recall are above 81\% and false
positive rate below 12\% across all services. The accuracy of the resolution inference
model is particularly high for YouTube and Twitch with average precision of 0.98 for both services.
The accuracy is slightly lower for Amazon and Netflix, because these services change the resolution
more frequently to adapt to playing conditions, whereas YouTube and Twitch often stick to a
single, often lower, resolution.

We also quantify the error rate for resolution inference.
Figure~\ref{fig:resolution_rel_error} shows the kernel density estimation
obtained plotting relative errors for the time slots for which the model
infers an incorrect resolution. Recall that we model resolution using a
multi-class classifier, resulting in discrete resolution ``steps''. Hence, the error
corresponds to distance in terms of the number of steps away from the correct
resolution. As with the startup delay, errors tend to be centered around the
real value, with the vast majority of errors falling within one or two
resolution steps away from the ground truth. For example, Netflix is the service
with the higher number of incorrectly inferred time slots at 14\%, but 83.9\%
of these errors are only one step away. Even given these error rates, this
model still offers a better approximation of actual video resolution than
previous models that only reflect coarse quality (\eg, good vs. bad).

\begin{figure}[t!]
\centering
\begin{minipage}{1\linewidth}
\centering
\begin{subfigure}[b]{0.95\linewidth}
\includegraphics[width=\linewidth]{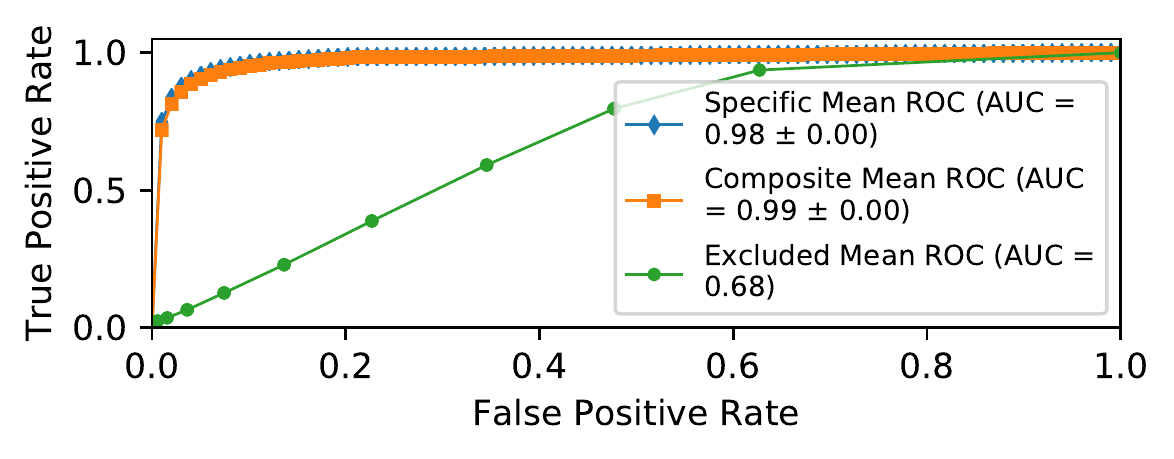}
\caption{ROC.}
\label{fig:composite_roc}
\end{subfigure} \hfill
\begin{subfigure}[b]{0.95\linewidth}
\includegraphics[width=\linewidth]{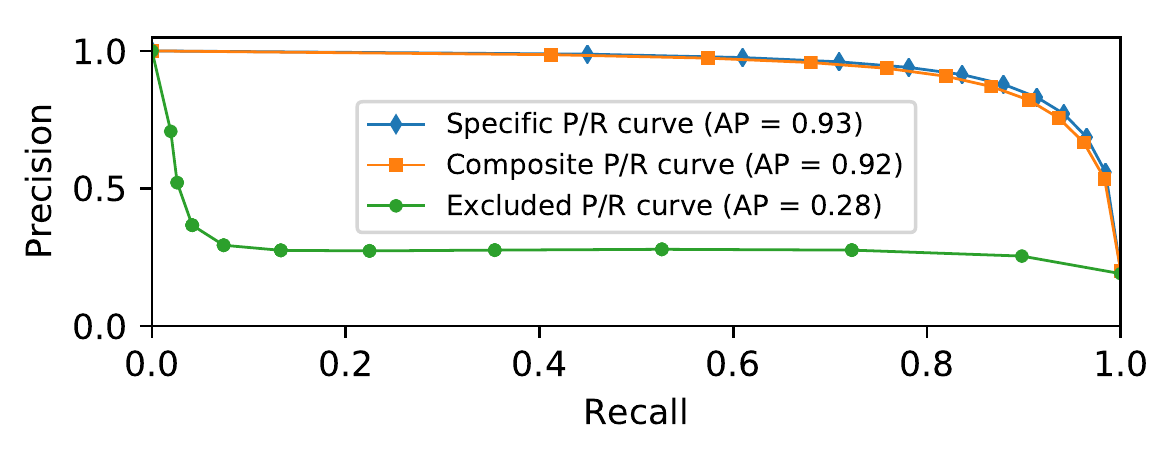}
\caption{Precision-recall.}
\label{fig:composite_pr}
\end{subfigure}
\end{minipage}
\caption{Composite vs. specific Net+App models performance for resolution inference.}
\label{fig:service_res}
\end{figure}

\paragraph{Composite models.} Figure~\ref{fig:service_res} shows model accuracy when
testing with Netflix using the specific model, the composite model, and the  excluded model.
We illustrate with results
for Netflix, but the conclusions are similar when
doing this analysis for the other three services. The composite model performs
nearly as well as the specific model (with average precision 0.92 compared to
0.93 for the specific model). These results are aligned with the general correlation
between segment sizes and resolution that we observed for all services.
Indeed, the four most important features for the composite model  are all
related to segment size: weighted average and standard deviation of
the sizes of the last ten downloaded segments, and the maximum and
average segment sizes. This observation thus provides a solid basis for
a composite model based on features related to segments.

\begin{figure}[t!]
\begin{minipage}{1\linewidth}
\centering
\begin{subfigure}[b]{0.47\linewidth}
\includegraphics[width=\linewidth]{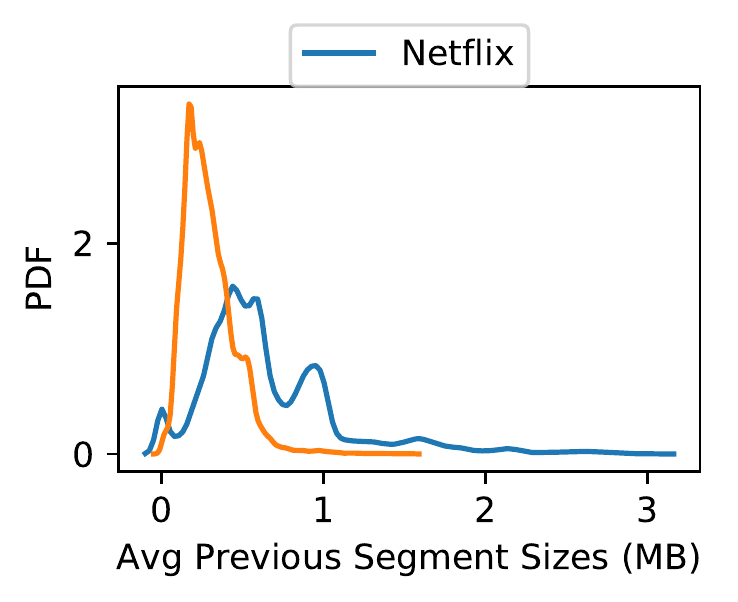}
\caption{Average previous segment sizes.}
\label{fig:avgsize}
\end{subfigure} \hfill
\begin{subfigure}[b]{0.47\linewidth}
\includegraphics[width=\linewidth]{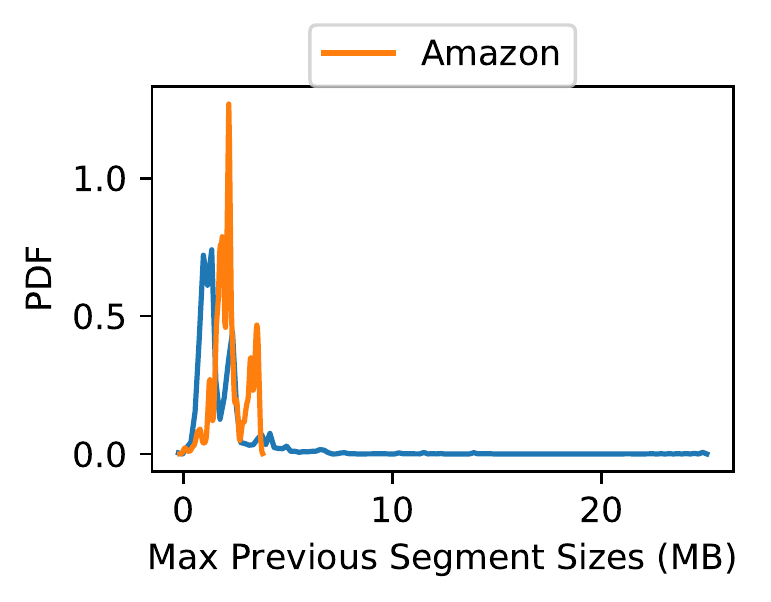}
\caption{Max size previous segments.}
\label{fig:maxsize}
\end{subfigure}
\end{minipage}
\caption{Both max previous segment size and average previous segment size differ across services.}
\label{fig:features_comp_res}
\end{figure}

\paragraph{Applying a composite model to new services.}
Figure~\ref{fig:service_res} also presents the accuracy of the excluded model, which is trained with data
from Amazon, YouTube, and Twitch and tested with
Netflix. As with the startup delay model, we observe a strong degradation of model
accuracy in this case. The average precision is only 0.2. Our analysis of the distributions of values
of the most important input features for the different services reveals even worse trends than discussed
for the startup delay. Although Netflix's behavior is most similar to Amazon's, there are sharp
differences across all the most important features. For example, Figure~\ref{fig:features_comp_res}
shows the  distribution for the two most important  features for Netflix, the average segment size of
previously downloaded segments and their max size. As a result, the accuracy of the
excluded model suffers. Our comprehensive analysis of the distributions of feature values across all
services indicates that achieving a general model for resolution seems even more challenging than for
startup delay as services' behavior vary substantially. As for startup delay, the rest of this paper will rely on
the composite model with Net+App features to infer resolution across services.

\section{Video Quality Inference in Practice}\label{sec:results}
We apply our models to network traffic collected from a year-long data
collection effort in 66 homes. In this section we present the challenges
we discovered when applying the methods in practice and techniques we
apply to mitigate them. We then use the Net+App, unified models from Section~\ref{sec:model-def}
to infer startup delay and resolution  on all video
sessions collected throughout the lifetime of the deployment
(Table~\ref{table:deployment_description}).

\subsection{In-Home Deployment Dataset}\label{sec:deployment}

\paragraph{Embedded home network monitoring system.}
We have developed a network monitoring system that collects the features shown
in Table~\ref{tab:features}. Our current deployment has been tested
extensively on both Raspberry Pi and Odroid platforms connected within
volunteers' homes. The system is flexible and can operate in a variety of
configurations: our current setup involves either setting the home gateway
into bridge mode and placing the device in between the gateway and the user's
access point, or deploying the device as a passive monitoring device on a span
port on the user's home network gateway. We present
the software architecture and design in a separate paper currently under
submission~\cite{bronzino2019network}.

The system collects network and application features aggregated across
five-second intervals. For each interval, we report average statistics for
network features divided per flow, together with the list of downloaded video
segments. Every hour, the system uploads the collected statistics to a remote
server. To validate the results produced by the model, we instrument the
extension presented in Section~\ref{sec:trainingdata} in five homes. Through
the extension we collect ground
truth data for 2,347 streaming sessions for the four services used to train
the models.

%We filter out all video
%sessions with startup delay larger than 60~seconds, as these sessions most
%likely correspond to instances when the video never played. The total number
%of discarded sessions is 4.7\% of the dataset.

\paragraph{Video sessions from 66 homes.} We analyze data
collected between January 23, 2018, and March 12, 2019. We concluded the data collection in May 2019, at which point we
had 60 devices in homes in the United States participating in our study, and
an additional 6 devices deployed in France. Downstream throughputs from
these networks ranged from 18~Mbps to 1~Gbps.
During the duration of the deployment, we have recorded a total of 216,173
video sessions from four major video service providers: Netflix,
YouTube, Amazon, and Twitch. Table~\ref{table:deployment_description} presents
an overview of the deployment, including the total number of video sessions
collected for each video service provider grouped by the homes'
Internet speed (as per the users' contract with their ISP) and the number of unique device MAC addresses seen in the home networks.

Additionally, we periodically (four times per day) record the Internet capacity
using active throughput measurements (\eg, speed tests) from the embedded system.
We collect this information to understand relationships between access link
capacity and video QoE metrics.

\begin{table}[t]
\centering
\scalebox{0.868}{
\begin{tabular}{|l|r|r|r|r|r|r|}
\hline
Speed &Homes&Devices&\multicolumn{4}{c|}{\# Video Sessions}\\
\cline{4-7}
[mbps] & & & Netflix & YouTube & Amazon & Twitch \\
\hline
(0,50] & 20 & 329 & 21,774 & 65,448 & 2,612 & 2,584\\
\hline
(50,100] & 23 & 288 & 11,788 & 50,178 & 3,273 & 3,345\\
\hline
(100,500] & 19 & 277 & 13,201 & 38,691 & 2,030 & 197\\
\hline
(500,1000] & 4 & 38 & 523 & 442 & 86 & 1\\
\hline
\hline
Total & 66 & 932 & 47,286 & 154,759 & 8,001 & 6,127 \\
\hline
\end{tabular}}
\caption{Number of video sessions per speed tier.}
\label{table:deployment_description}
\end{table}

\begin{figure}[t]
\begin{center}
\includegraphics[width=0.75\linewidth,trim={0 0 0 0},clip]{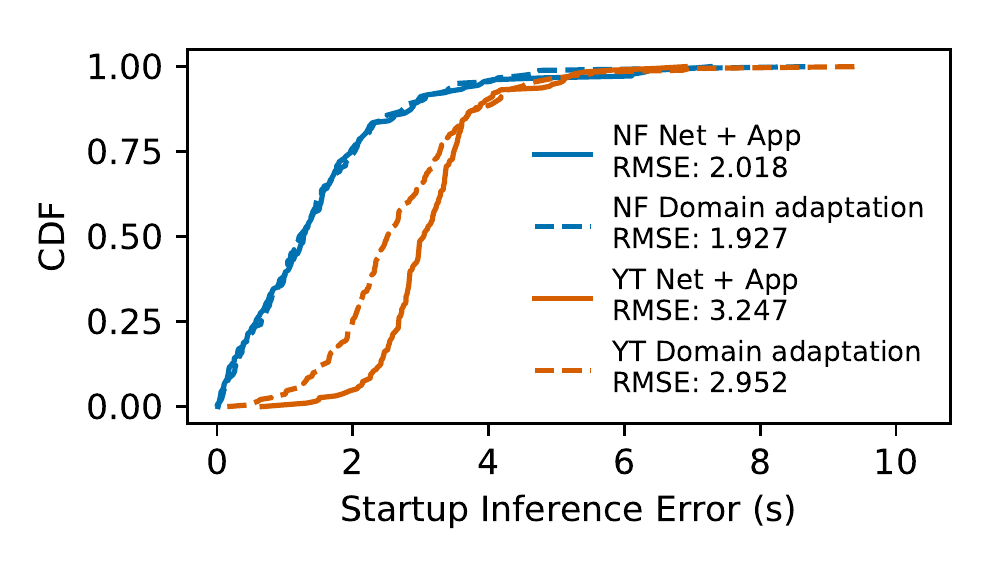}
\vspace{-3mm}
\caption{Effect of domain adaptation on startup delay inference.}
\label{fig:fuzzycompare}
\end{center}
% \vspace{-3mm}
\end{figure}

\subsection{Practical Challenges for Robust Models}\label{sec:robust-models} \label{sec:practical}
Testing our models
in a long-running deployment raises a new set of
challenges that are not faced by offline models that operate on curated traces in controlled lab
settings. Two factors, in particular, affected the accuracy of the models:
(1)~the granularity of training data versus what is practical to collect in
an operational system; and (2)~the challenge of accurately detecting the start and end of a
video session in the presence of unrelated cross-traffic.%, and
%attributing the video session traffic to a particular service.

\paragraph{Granularity and timing of traffic measurements.} An operational monitoring
system cannot export information about each individual packet. It is hence common
to report traffic statics in fixed time intervals or time bins (\eg, SNMP or Netflow polling
intervals).  The training data that we and prior work collect has a precise session start time,
whereas the data collected from a deployed system will only have data collected in time
intervals, where the session start time might be {\em anywhere} in that
interval. This corresponding mismatch in
granularity creates a challenge for training. Furthermore, any error in
estimating the start time propagates to the time bins used for inference
across the entire session, resulting in a situation where the time bins in the
training and deployment data sets do not correspond to one another at all.

\paragraph{Session detection}
Identifying  a video session
from encrypted network traffic is a second challenge as network traffic is
noisy. First, one must identify the relatively few flows carrying traffic of a
given video service out of all the haystack of flows traversing a typical
network monitor. Second, not all flows of a video service carries video. For
example, some flows may carry the elements that compose
the webpage where users can browse videos and others may carry ads. Finally,
within the video flows, one must be able to identify when each video session
starts and ends.

To detect session start and end times, we extend the method from
Dimopoulos~\ea~\cite{dimopoulos2016measuring},
which identifies a spike in traffic to specific YouTube domains to determine
the start of a video session and a silent period to indicate the end of a
video session. We evaluate their method with our ground truth and find
that the session start time error
typically falls between zero and five seconds due to the granularity of the
monitoring system, but in some
cases---because browsing a video catalog sometimes initiates the playback of a
movie trailer, for example---the method may incorrectly inflate the length of
any particular video session. We present further validation
in~\cite{bronzino2019network}.

%\begin{figure*}[t!]
%\begin{subfigure}[t]{0.24\linewidth}
%\includegraphics[width=\linewidth]{figures/startup_vs_tier_bismark_Netflix.pdf}
%\caption{Netflix.}
%\label{fig:startup_bismark_net}
%\end{subfigure}%
%~
%\begin{subfigure}[t]{0.24\linewidth}
%\includegraphics[width=\linewidth]{figures/startup_vs_tier_bismark_Youtube.pdf}
%\caption{YouTube.}
%\label{fig:startup_bismark_you}
%\end{subfigure}
%~
%\begin{subfigure}[t]{0.24\linewidth}
%\includegraphics[width=\linewidth]{figures/startup_vs_tier_bismark_Amazon.pdf}
%\caption{Amazon.}
%\label{fig:startup_bismark_amz}
%\end{subfigure}
%~
%\begin{subfigure}[t]{0.24\linewidth}
%\includegraphics[width=\linewidth]{figures/startup_vs_tier_bismark_Twitch.pdf}
%\caption{Twitch.}
%\label{fig:startup_bismark_twi}
%\end{subfigure}
%\caption{Startup Delay Inference vs. Active Throughput Measurements (95th
%Percentile).}
%\label{fig:startup_bismark}
%\end{figure*}

\begin{table}[t]
% \Small
\centering

\scalebox{0.95}{
\begin{tabular}{|c|c|c|c|c|c|c|}
\cline{2-7}
\multicolumn{1}{c|}{}&\multicolumn{3}{c|}{Net + App} &\multicolumn{3}{c|}{Domain adaptation} \\
\cline{2-7}
\multicolumn{1}{c|}{}&Precision & Recall & FPR & Precision & Recall & FPR \\
\hline Netflix & 86\% & 86\% & 3.4\% & 90\% & 90\% & 2.5\%\\
\hline YouTube & 82\% & 82\% & 4.4\% & 78\% & 78\% & 5.5\%\\
\hline Twitch & 82\% & 82\% & 4.6\% & 91\% & 91\% & 2.2\%\\
\hline Amazon & 76\% & 76\% & 6\% & 78\% & 78\% & 5.6\%\\
\hline
\end{tabular}}
\caption{Effect of domain adaptation on resolution inference.}
\label{table:fuzzy_compare_resolution}
% \vspace{-3mm}
\end{table}

\begin{figure}[t]
\begin{center}
\vspace{-3mm}
\includegraphics[width=0.9\linewidth]{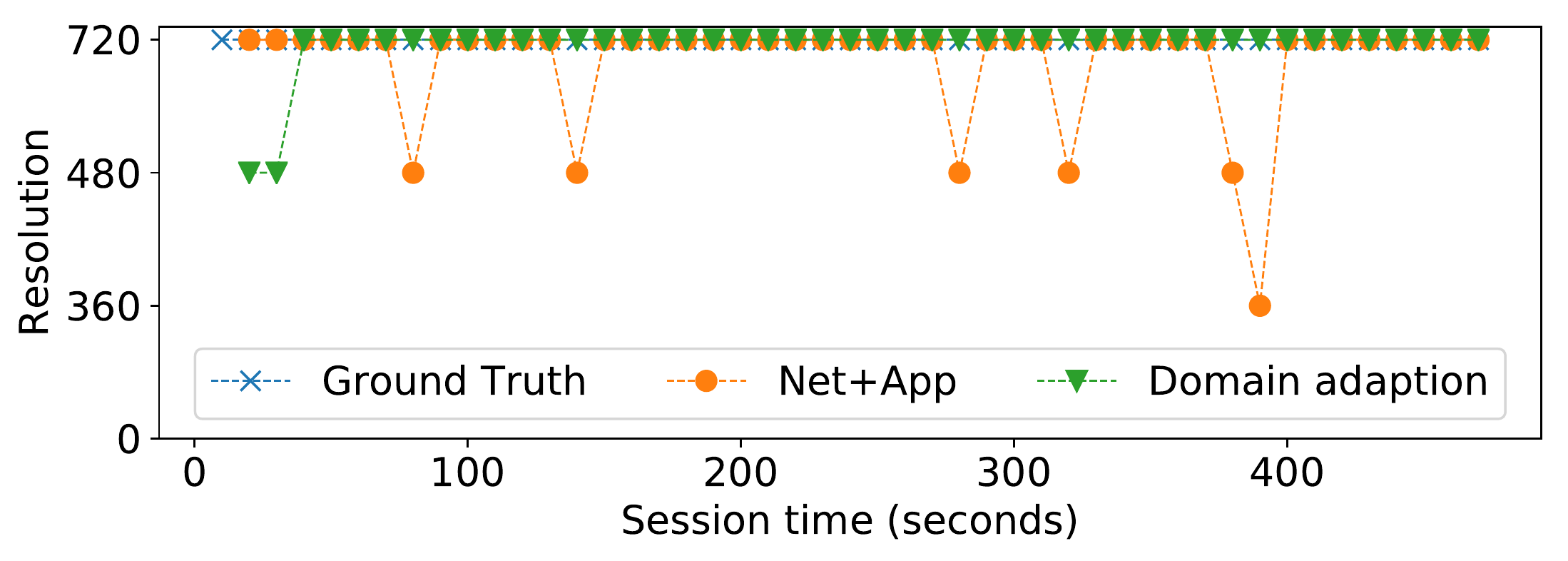}
\caption{Sample session for errors with/without domain adaptation.}
\label{fig:fuzzyresolutionswitch}
\end{center}
\vspace{-3mm}
\end{figure}

\begin{figure*}[t!]
\begin{subfigure}[t]{0.24\linewidth}
\includegraphics[width=\linewidth]{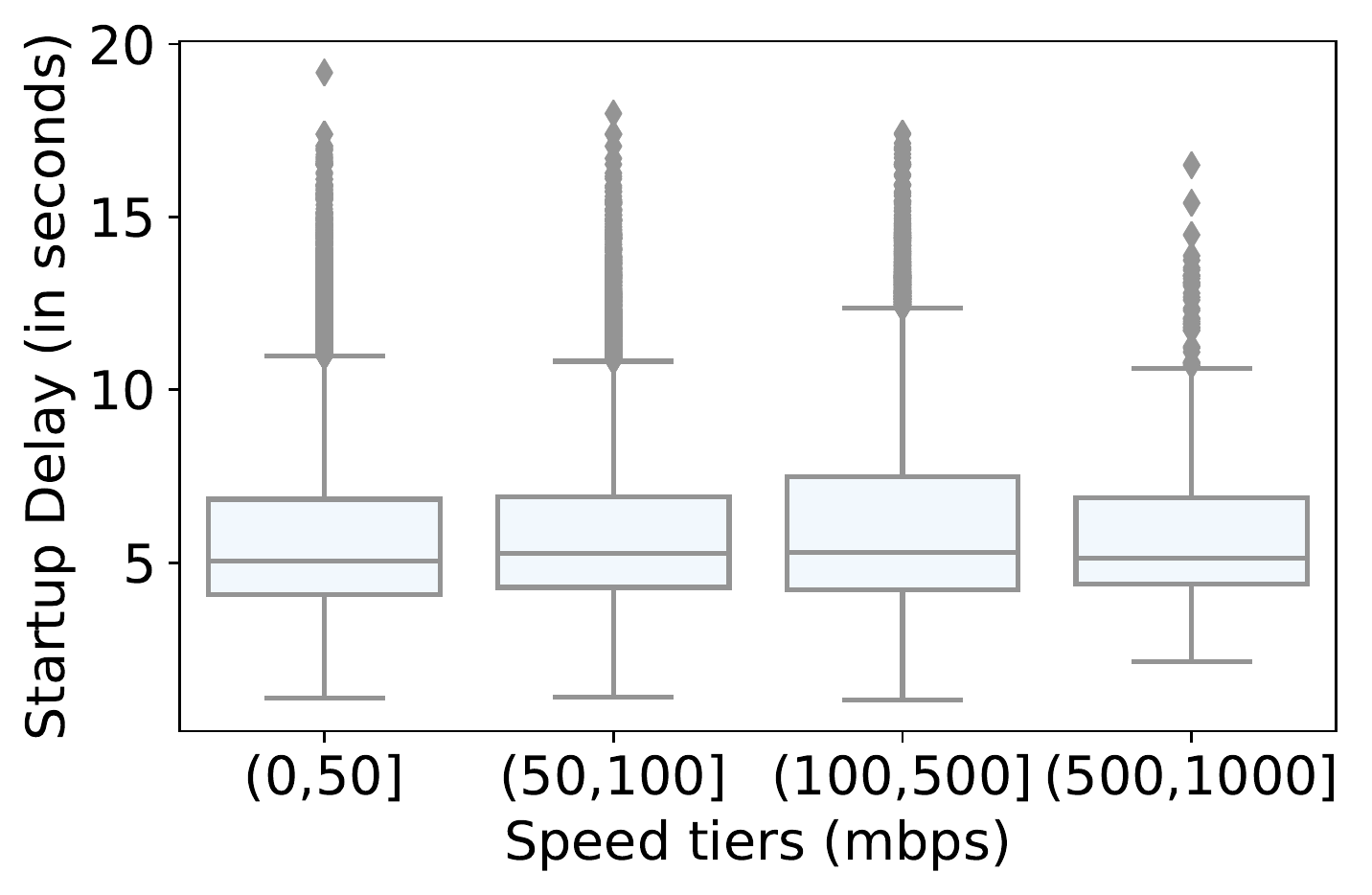}
\caption{Netflix.}
\label{fig:startup_nominal_net}
\end{subfigure}%
~
\begin{subfigure}[t]{0.24\linewidth}
\includegraphics[width=\linewidth]{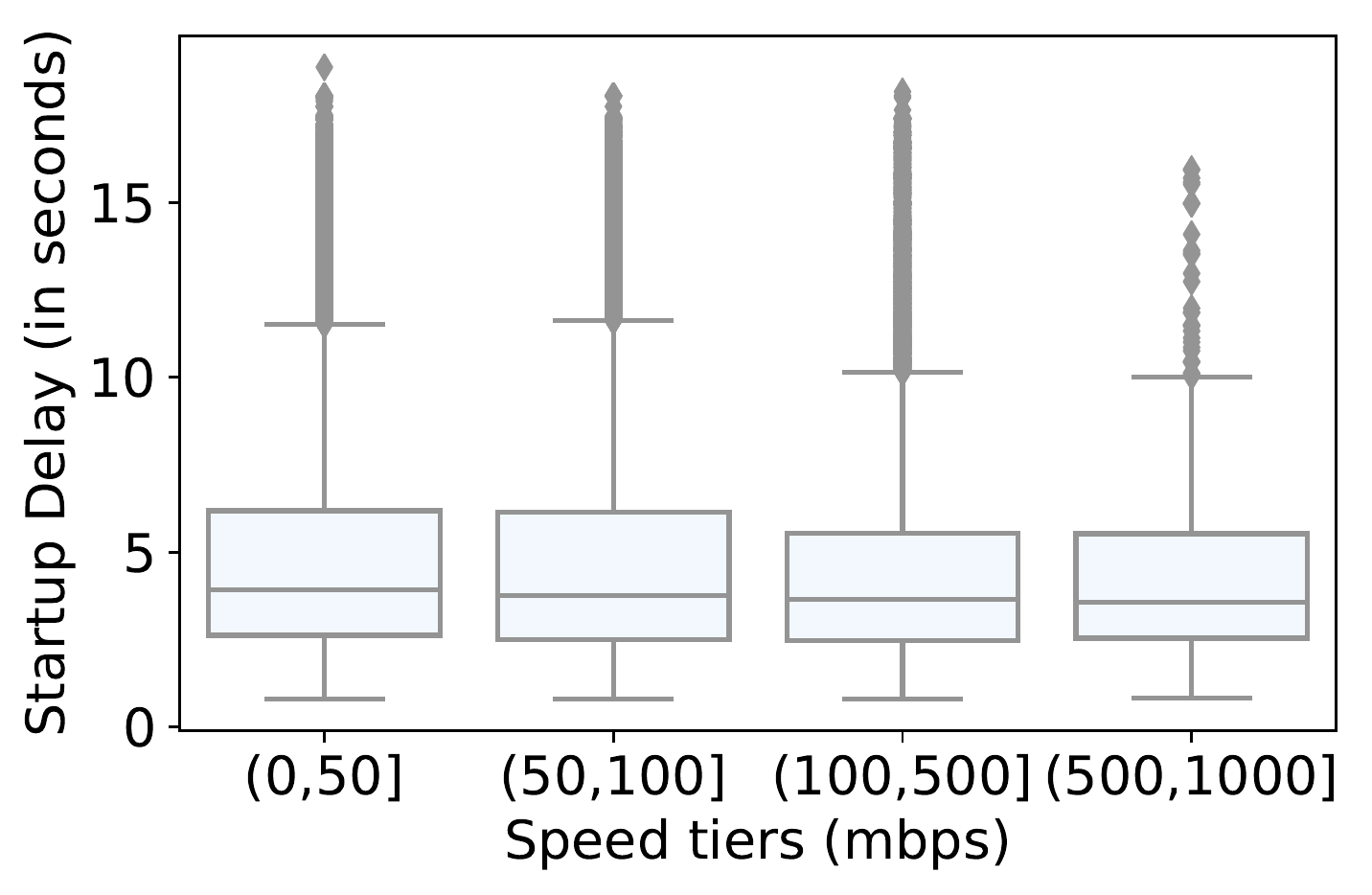}
\caption{YouTube.}
\label{fig:startup_nominal_you}
\end{subfigure}
~
\begin{subfigure}[t]{0.24\linewidth}
\includegraphics[width=\linewidth]{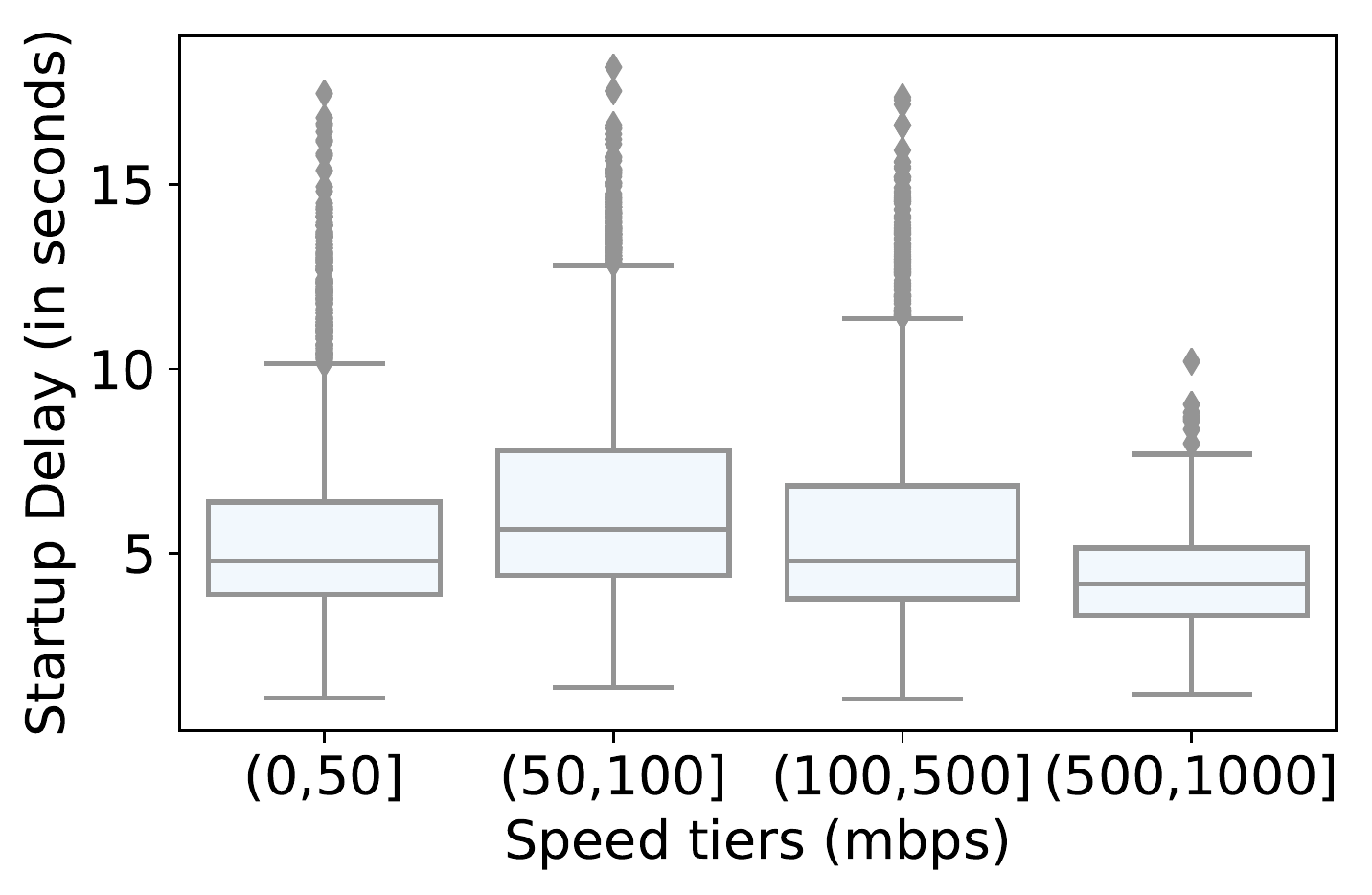}
\caption{Amazon.}
\label{fig:startup_nominal_amz}
\end{subfigure}
~
\begin{subfigure}[t]{0.24\linewidth}
\includegraphics[width=\linewidth]{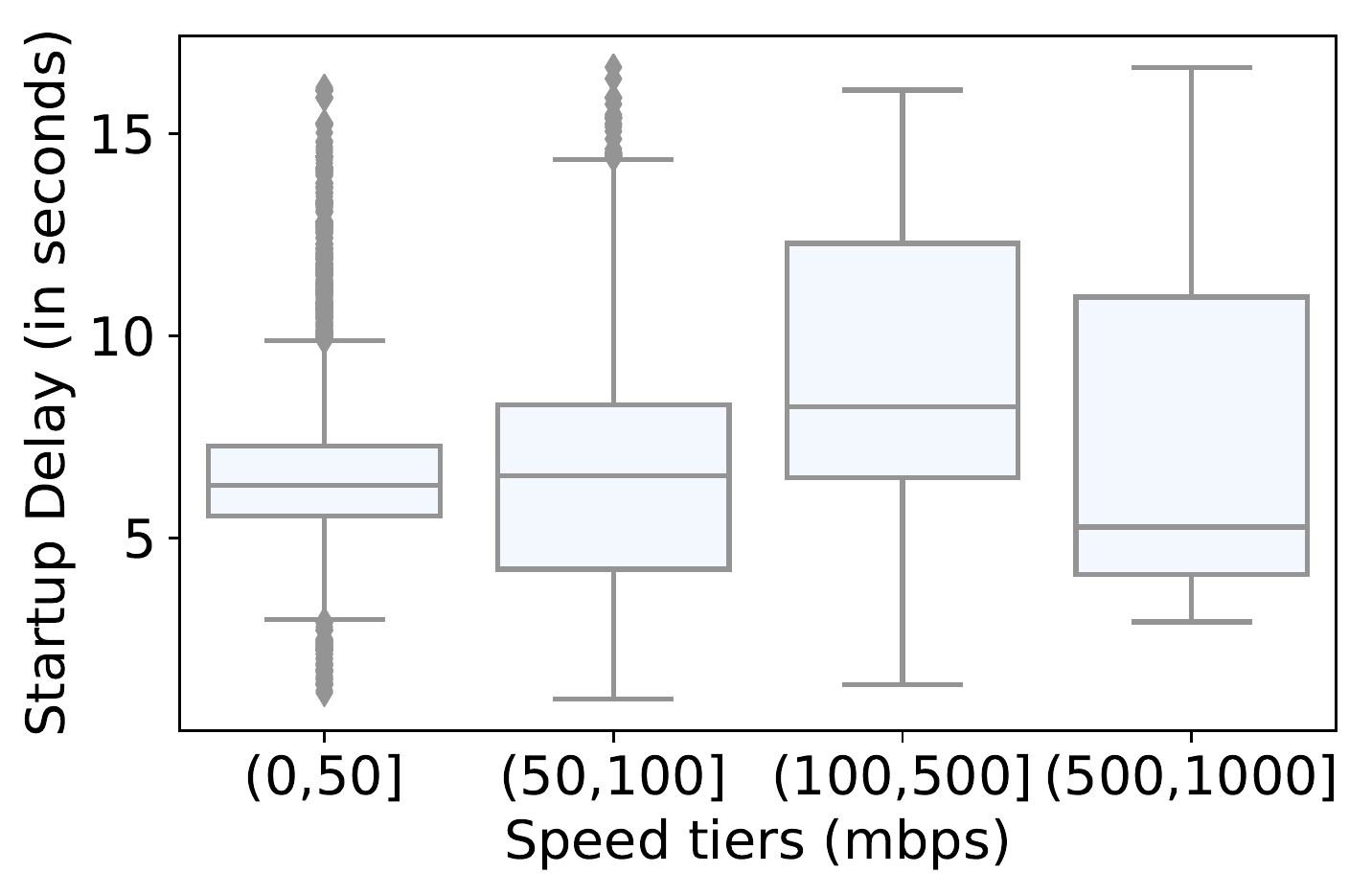}
\caption{Twitch.}
\label{fig:startup_nominal_twi}
\end{subfigure}
\vspace{-3mm}
\caption{Startup Delay Inference vs. Nominal Speed Tier.}
\label{fig:startup_nominal}
\vspace{-3mm}
\end{figure*}

\begin{figure*}[t!]
\begin{subfigure}[t]{0.24\linewidth}
\includegraphics[width=\linewidth]{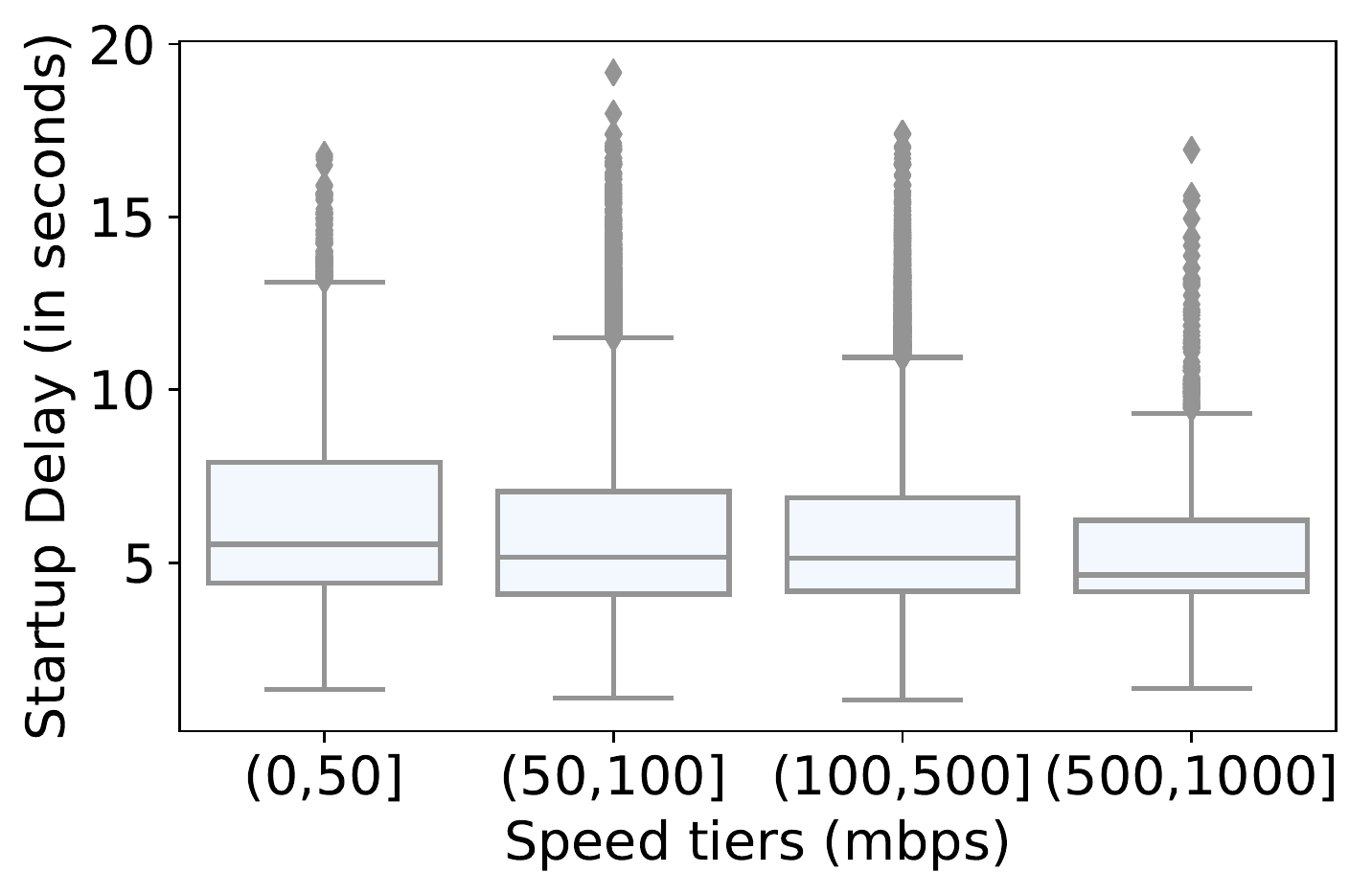}
\caption{Netflix.}
\label{fig:startup_bismark_net}
\end{subfigure}%
~
\begin{subfigure}[t]{0.24\linewidth}
\includegraphics[width=\linewidth]{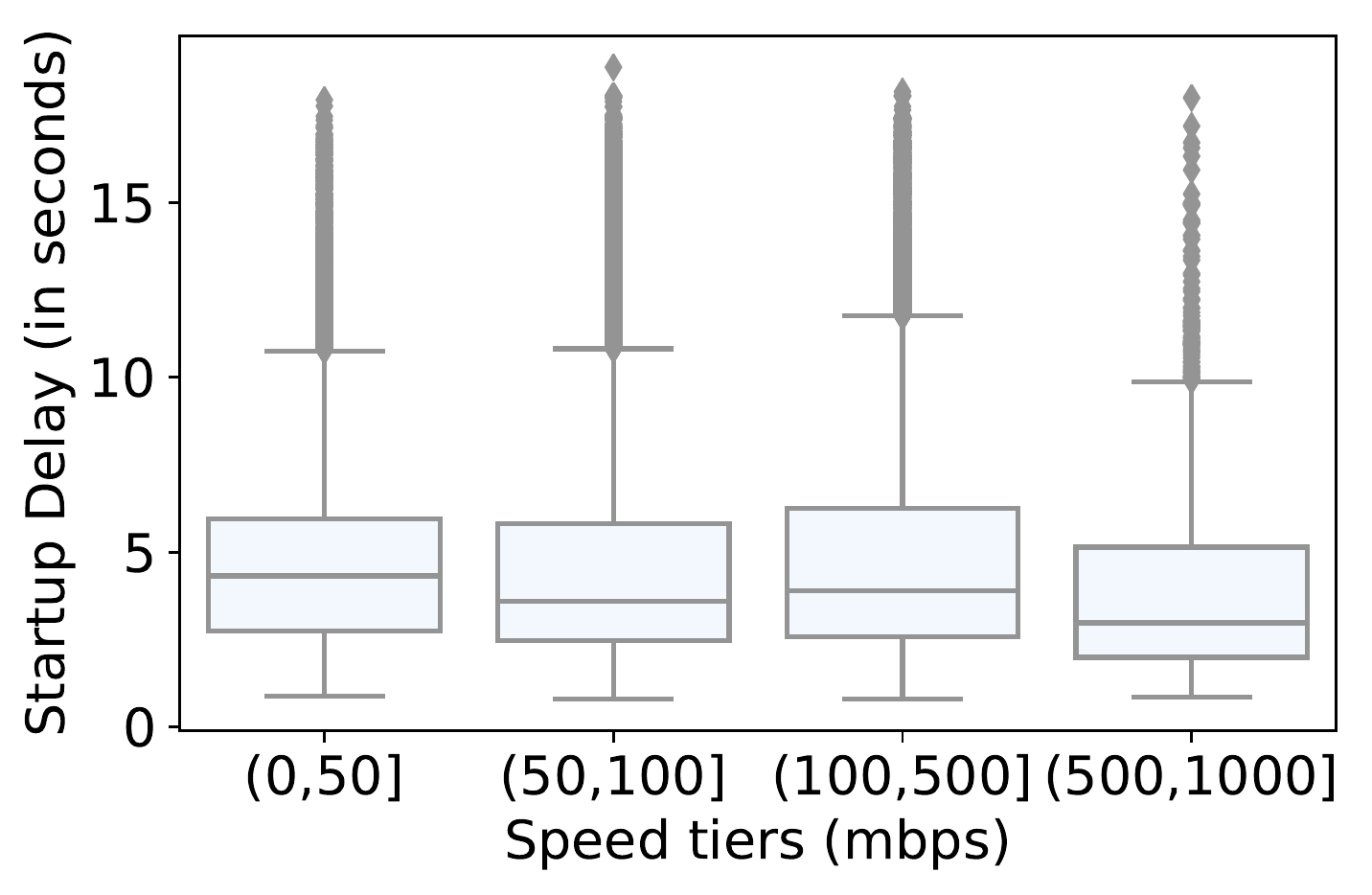}
\caption{YouTube.}
\label{fig:startup_bismark_you}
\end{subfigure}
~
\begin{subfigure}[t]{0.24\linewidth}
\includegraphics[width=\linewidth]{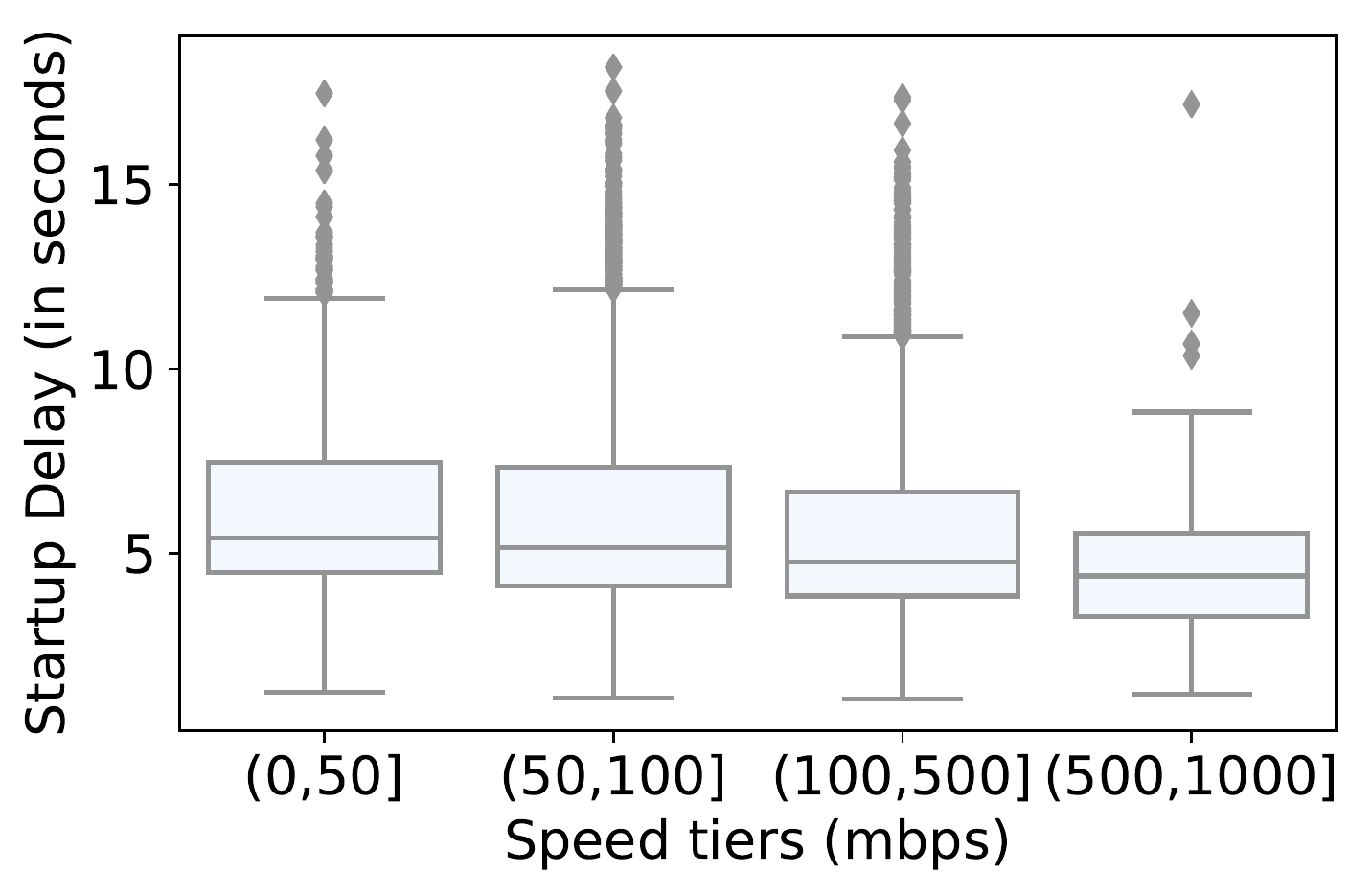}
\caption{Amazon.}
\label{fig:startup_bismark_amz}
\end{subfigure}
~
\begin{subfigure}[t]{0.24\linewidth}
\includegraphics[width=\linewidth]{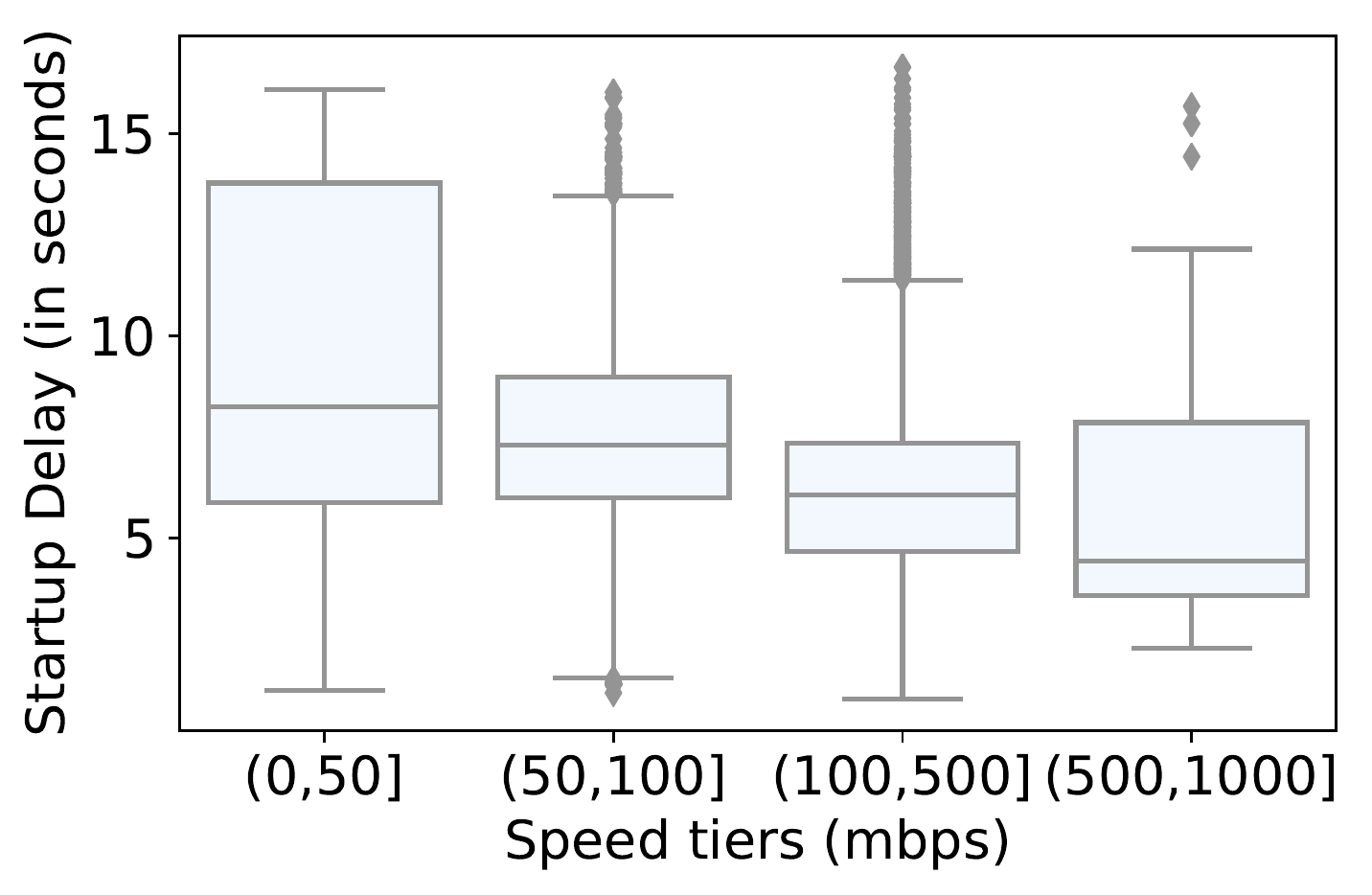}
\caption{Twitch.}
\label{fig:startup_bismark_twi}
\end{subfigure}
\vspace{-3mm}
\caption{Startup Delay Inference vs. Active Throughput Measurements (95th
Percentile).}
\label{fig:startup_bismark}
\vspace{-3mm}
\end{figure*}

\begin{figure*}[t!]
\begin{subfigure}[b]{0.24\linewidth}
\includegraphics[width=\linewidth]{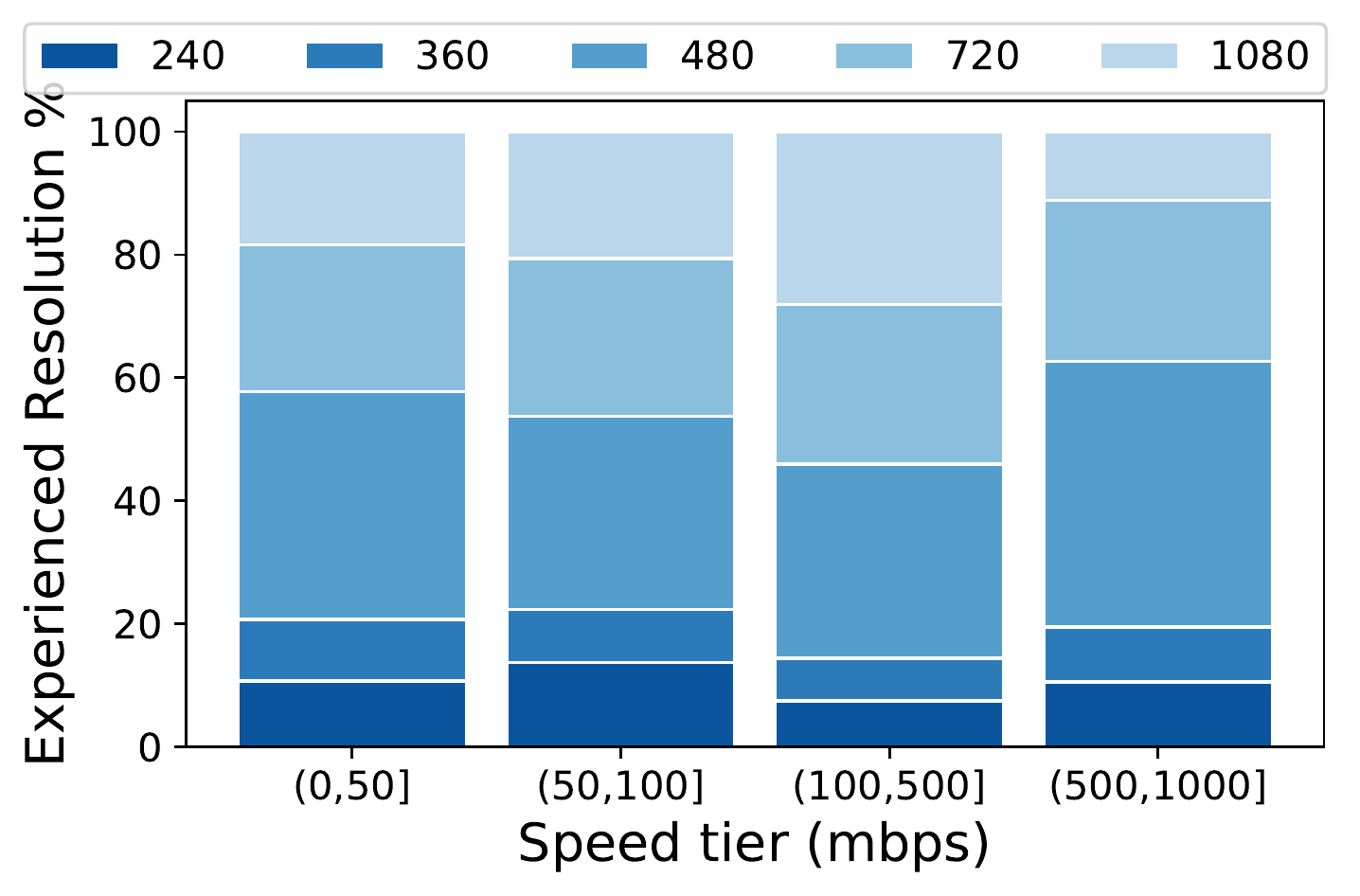}
\caption{Netflix.}
\label{fig:nominal_net}
\end{subfigure}
\hfill
\begin{subfigure}[b]{0.24\linewidth}
\includegraphics[width=\linewidth]{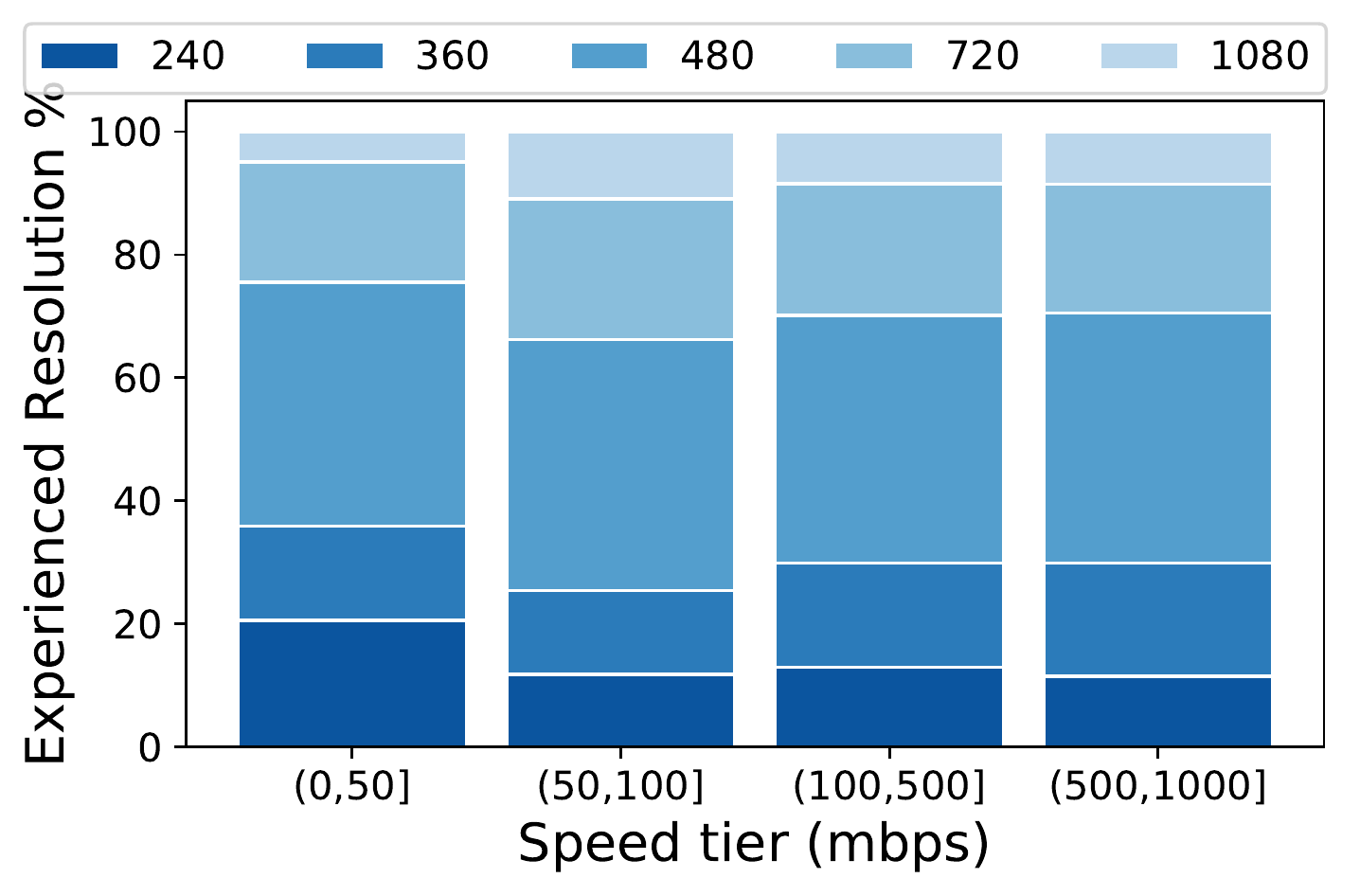}
\caption{YouTube.}
\label{fig:nominal_you}
\end{subfigure}
\hfill
\begin{subfigure}[b]{0.24\linewidth}
\includegraphics[width=\linewidth]{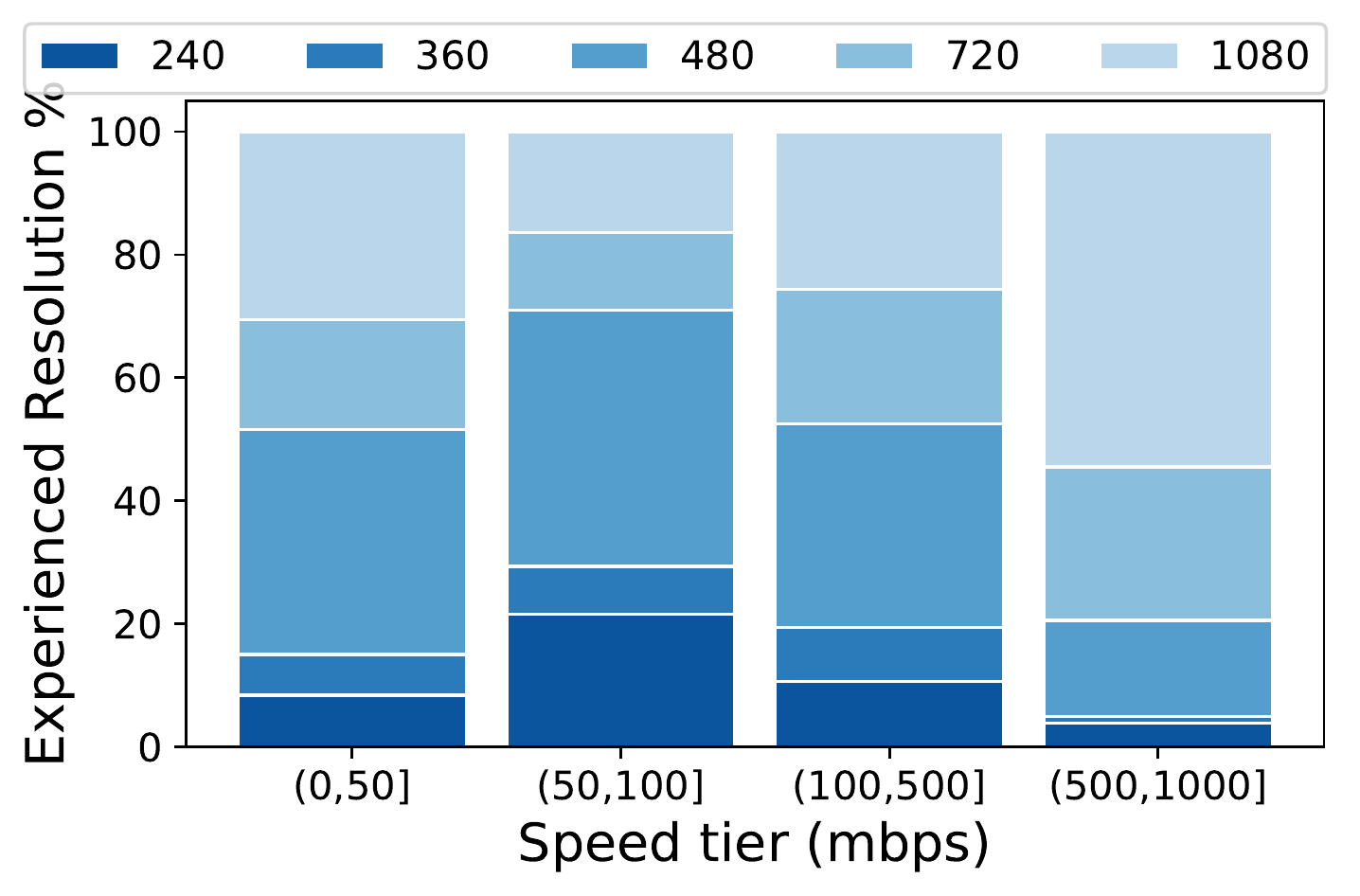}
\caption{Amazon.}
\label{fig:nominal_amz}
\end{subfigure}
\hfill
\begin{subfigure}[b]{0.24\linewidth}
\includegraphics[width=\linewidth]{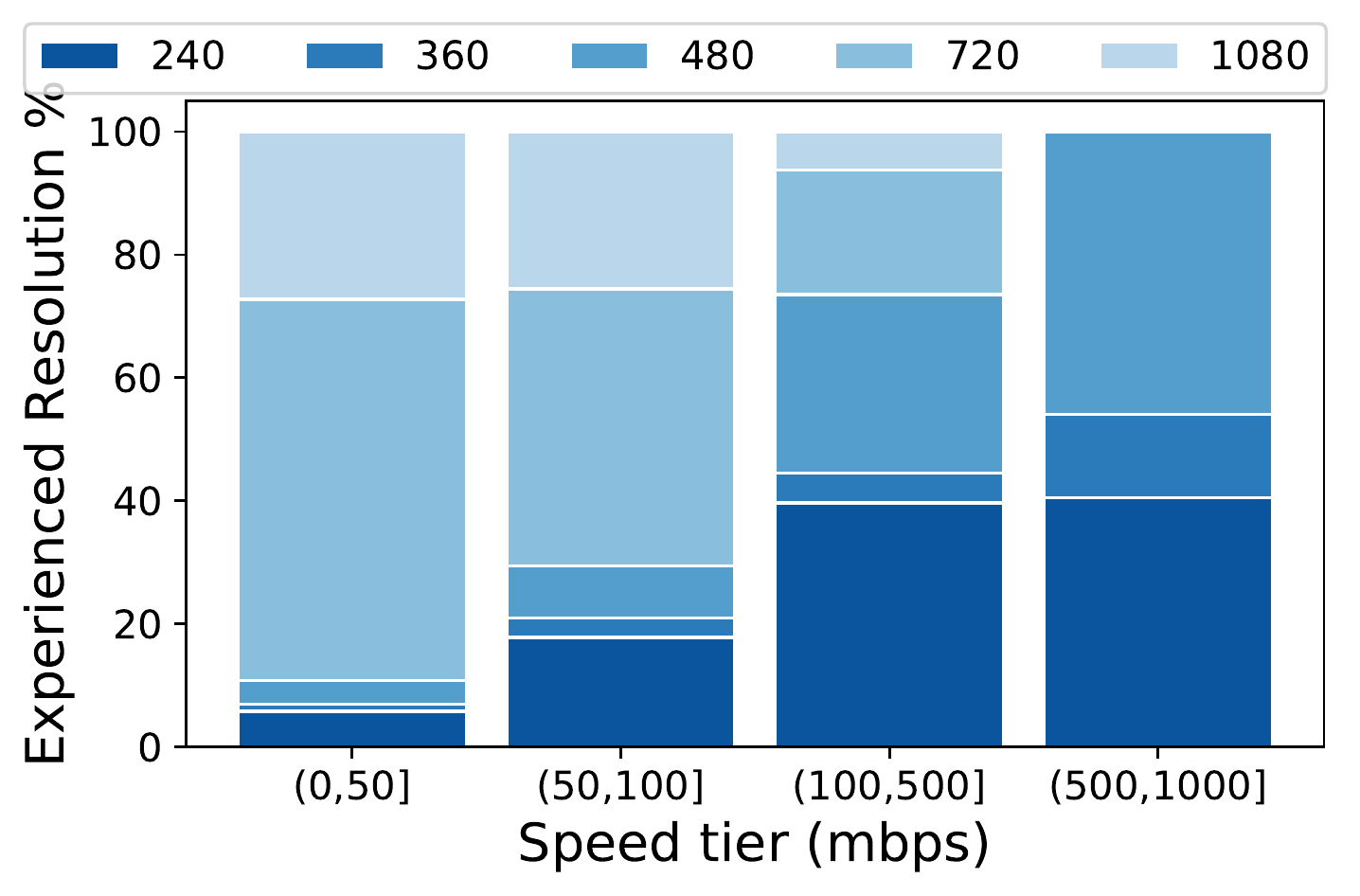}
\caption{Twitch.}
\label{fig:nominal_twi}
\end{subfigure}
\vspace{-3mm}
\caption{Resolution vs Nominal Speed Tier}
\label{fig:nominal}
\vspace{-3mm}
\end{figure*}

\begin{figure*}[t!]
\begin{subfigure}[b]{0.24\linewidth}
\includegraphics[width=\linewidth]{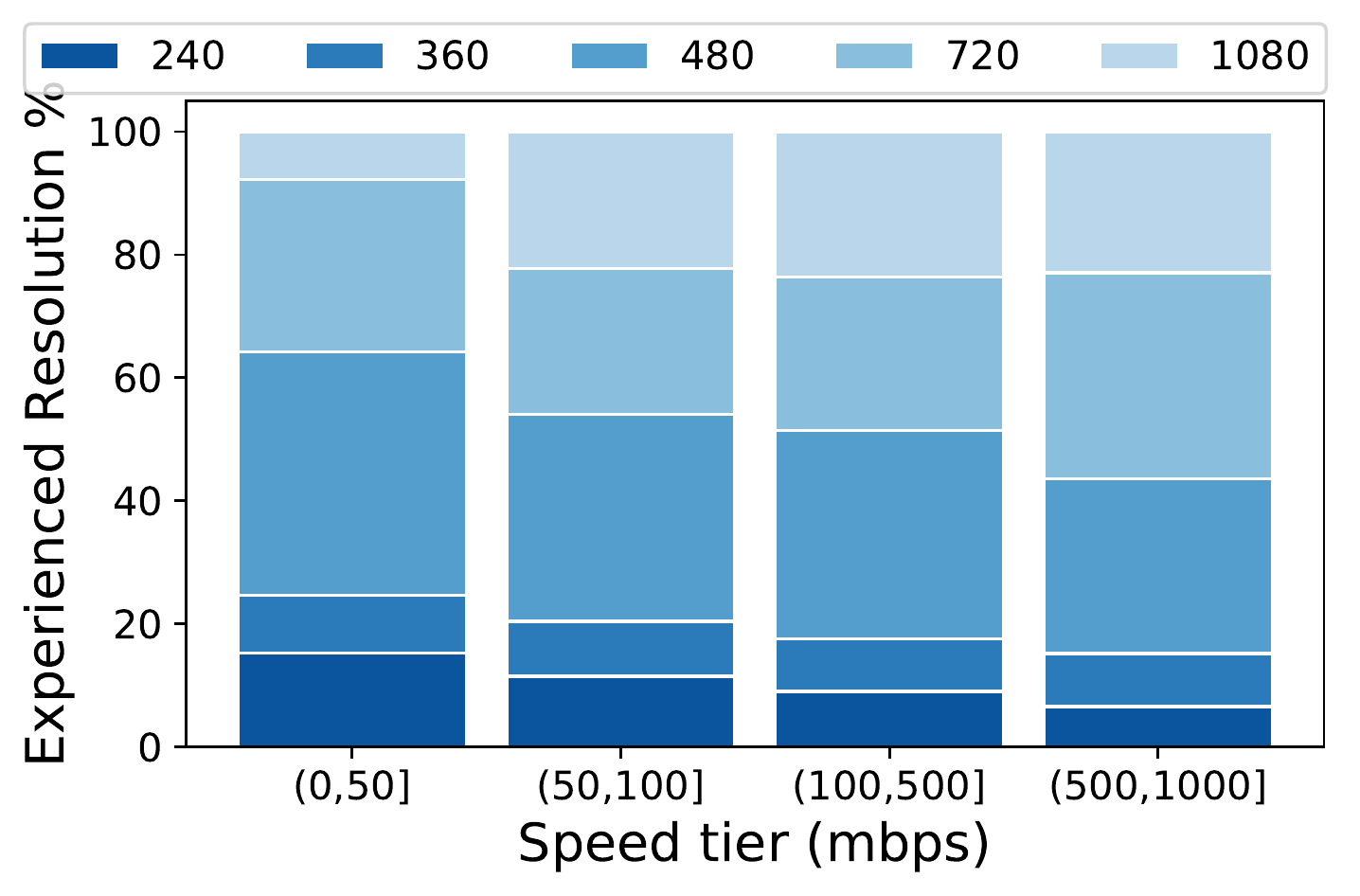}
\caption{Netflix.}
\label{fig:bismark_net}
\end{subfigure}
\hfill
\begin{subfigure}[b]{0.24\linewidth}
\includegraphics[width=\linewidth]{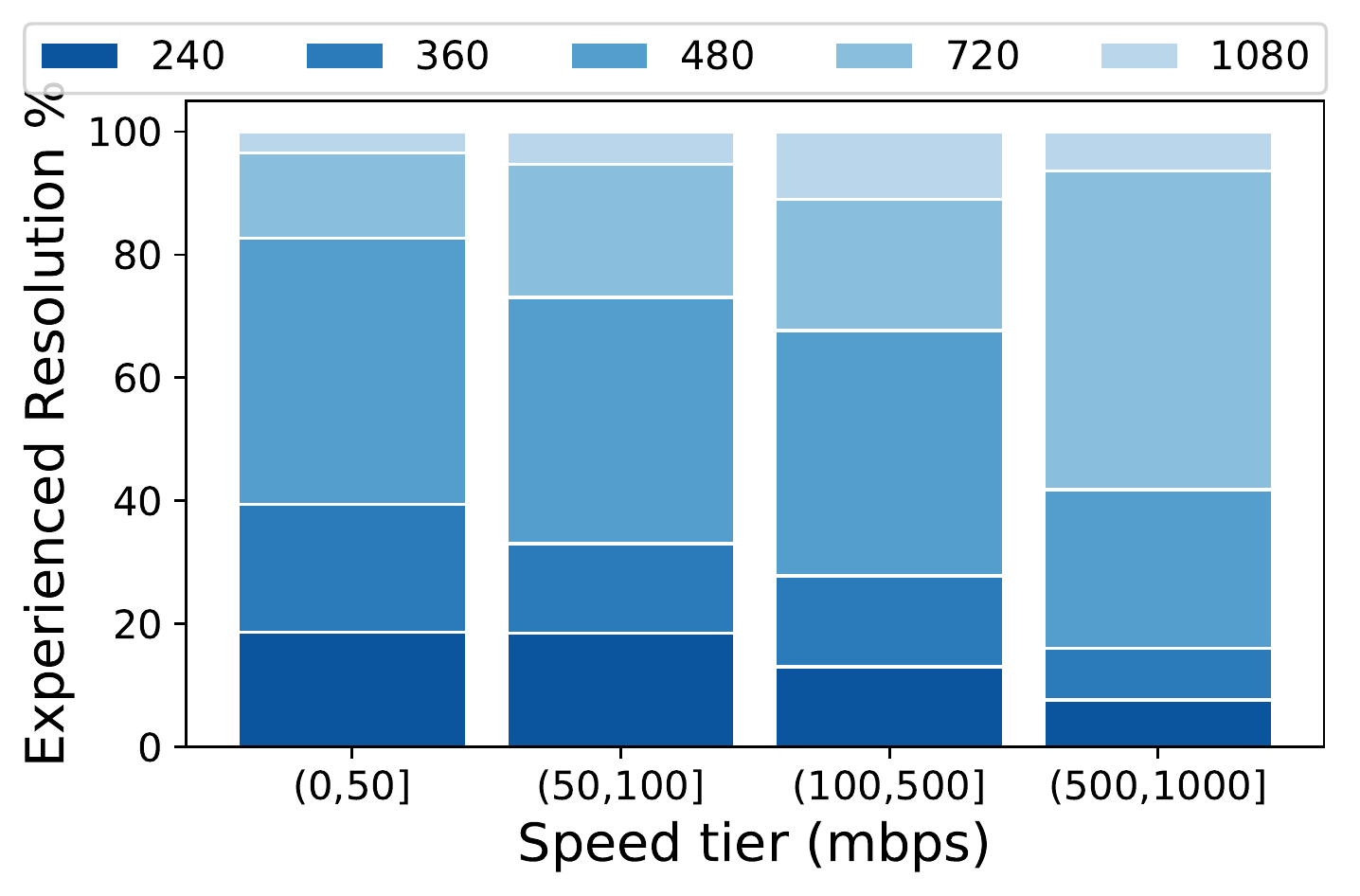}
\caption{YouTube.}
\label{fig:bismark_you}
\end{subfigure}
\hfill
\begin{subfigure}[b]{0.24\linewidth}
\includegraphics[width=\linewidth]{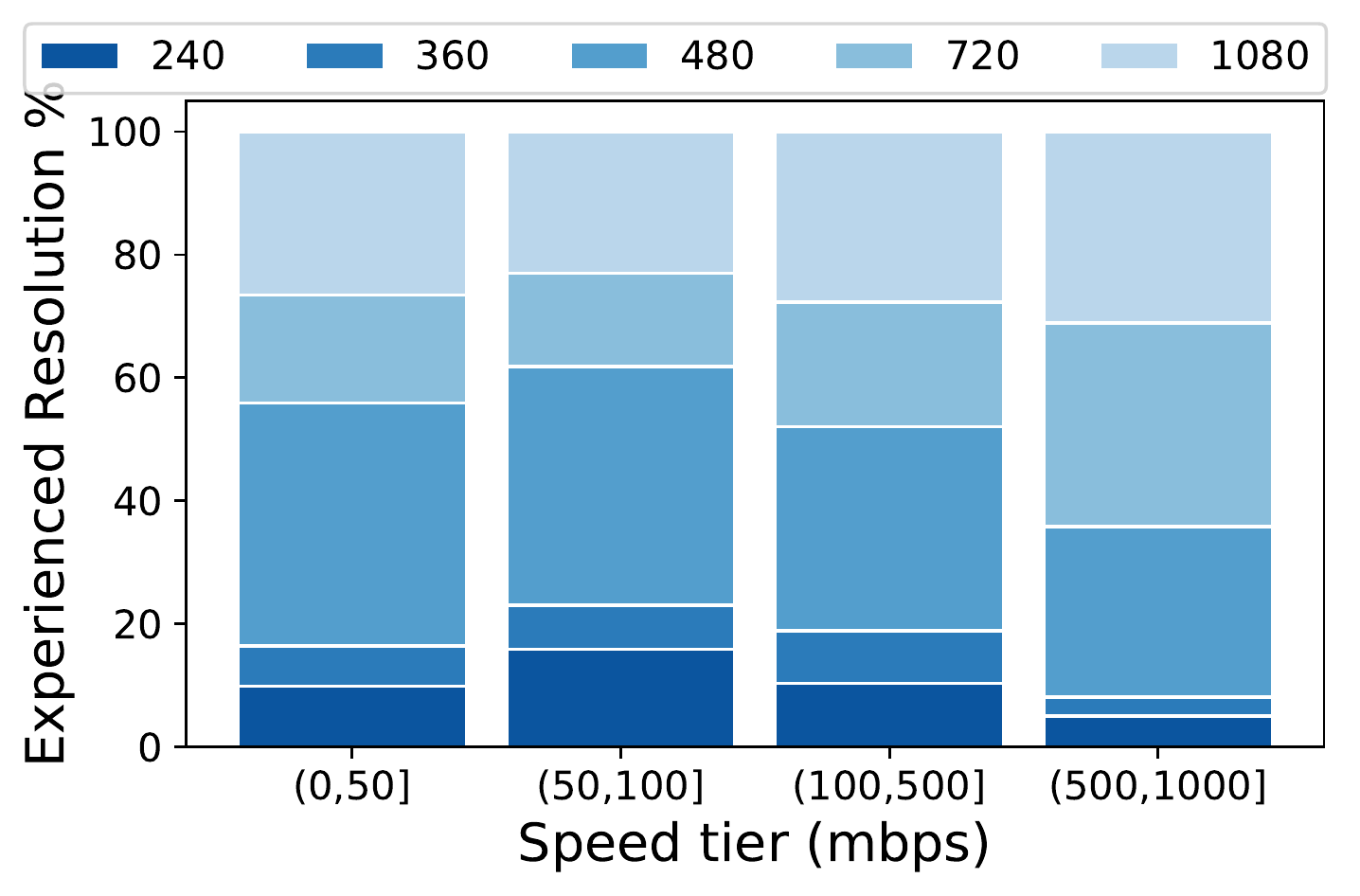}
\caption{Amazon.}
\label{fig:bismark_amz}
\end{subfigure}
\hfill
\begin{subfigure}[b]{0.24\linewidth}
\includegraphics[width=\linewidth]{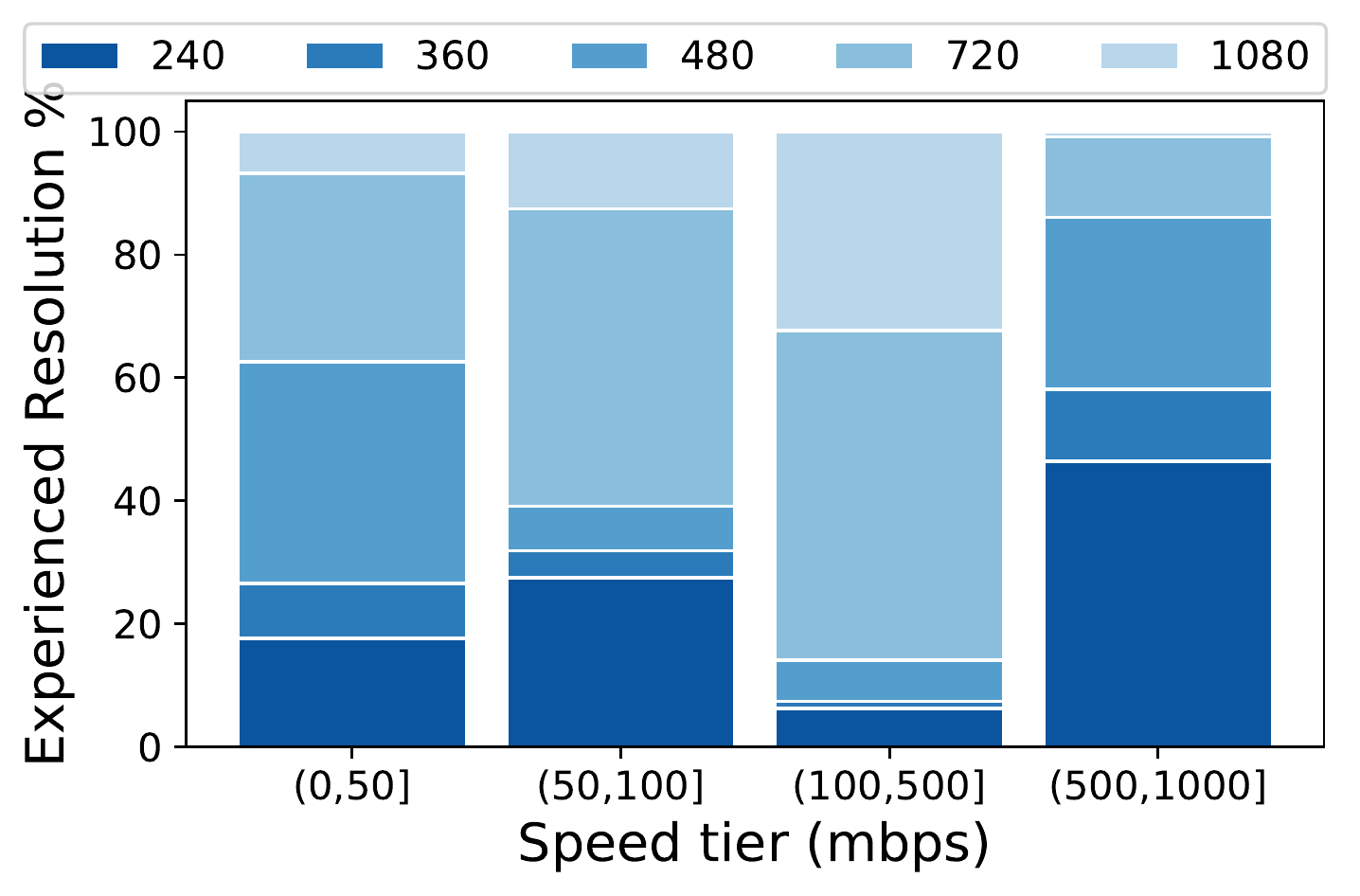}
\caption{Twitch.}
\label{fig:bismark_twi}
\end{subfigure}
\vspace{-3mm}
\caption{Resolution vs. Active Throughput Measurements (95th Percentile).}
\label{fig:bismark}
\vspace{-3mm}
\end{figure*}

\paragraph{Domain adaptation.} Intuitively, our approach to address these challenges involves accounting for additional noise that the
practical monitoring introduces that is not present in a lab setting or in the
training data. To do so, we introduce noise into the training data so that the
training data more closely resembles the data collected from deployment. The
techniques that we apply are grounded in the general theory of domain
adaptation~\cite{sun2016return}; they work as follows: Because the actual start
time can fall anywhere within the five-second interval, we pre-processed our
training data and artificially adjusted each session start time over a window of
-5 seconds to +5 seconds from the actual start value in increments of 0.5
seconds. For each new artificial start time, we recalculated all metrics based on this value for
the entire session. This technique has two benefits: it makes the model more
robust to noise, and it increases the volume of training data.

\paragraph{Validation.} We validated the domain-adapted models against 2,347
sessions collected across five homes of the deployment. We compared each
session collected from the extension with the closest session detected in the
deployment data. We present results for both startup delay and resolution
using the Net+App, unified model (as presented in Section~\ref{sec:model-def}) and the
model after domain adaptation.

Figure~\ref{fig:fuzzycompare} shows the resulting improvement in startup
inference when we apply domain adaption based models to Netflix and YouTube:
In both cases, the root mean square error improves---quite significantly for
YouTube. Table~\ref{table:fuzzy_compare_resolution} shows that we obtain
similar improvements for resolution inference, except for YouTube. We posit
that this result is attributable to the ground truth dataset collected; the YouTube
data is heavily biased towards 360p resolutions (90+\%), whereas all other
services operated at higher (and more diverse) resolutions. While domain
adaptation increases balance across classes, it may slightly
impact classes with more prevalence in the training dataset. We leave
investigation into mitigating such issues for future exploration. Finally, Figure~\ref{fig:fuzzyresolutionswitch} illustrates how domain adaption helps
correct errors on resolution inference through an example session.
These results suggest that domain adaptation is a promising approach for bridging
the gap between lab-trained models and real network deployments.  We expect
that results could be further improved by applying domain adaptation with smaller time
intervals.

\subsection{Inference Results}
We ran the quality inference models from Section \ref{sec:model-def} on all video
sessions collected throughout the lifetime of the deployment
(Table~\ref{table:deployment_description}).

\subsubsection{Startup Delay}
We infer the startup delay for each video session in the deployment dataset in
order to pose a question that both ISPs and customers may ask: how does access
link capacity relate to startup delay? Answering this question allows us to
understand the tangible benefits of paying for higher access capacity with
regard to video streaming. We employ two metrics of ``capacity'': the nominal
download subscription speed that users reported for each home, and the 95th
percentile of active throughput measurements that we collected
from the embedded in-home devices. Figure~\ref{fig:startup_nominal} presents box
plots of startup delay versus nominal speeds, whereas
Figure~\ref{fig:startup_bismark} presents startup delay versus the 95th
percentile of throughput measurements. Note that the number of samples for the
highest tier for both Amazon and Twitch is too small to draw conclusions
see Table~\ref{table:deployment_description}), so we focus on results in the
first three tiers for these services.

Figure~\ref{fig:startup_nominal} shows our unexpected result. We see that median
startup delays for each service tend to be similar across the subscription tiers.
For example, YouTube consistently achieves a median of \textless 5~second startup
delay across all tiers and Amazon a constant median startup delay of \textless
6~seconds. Netflix and Twitch's startup delay differs by at most $\pm$ 2~seconds
across tiers, but these plots exhibit no trend of decreasing startup delay as
nominal speeds increases as one would expect. There are several possible
explanations for this result. One possibility is that actual speeds vary
considerably over time due to variations in available capacity along end-to-end
paths due (for example) to diurnal traffic demands.

Startup delays follow a more expected pattern when we consider capacity as
defined by active throughput measurements. Figure~\ref{fig:startup_bismark}
shows that startup delay decreases as measured throughput increases for Netflix,
YouTube, and Twitch. For Netflix, the difference in startup delay between the
highest and lowest tier is four seconds, this difference is three seconds for
YouTube. This figure also highlights the difference in startup delay across
services, given similar network conditions. These differences reflect the fact
that different services select different design tradeoffs that best fit their
business model and content. For example, Netflix has the highest startup delay
of all the services. Because Netflix content is mostly movies and long-form
television shows---which are relatively long---having higher startup delays may
be more acceptable. As we see in the next section, our inference reflects the
expected  trends that Netflix trades off higher startup delay to achieve higher
resolution video streams.

\subsubsection{Resolution}
Next, we infer the resolution for each ten-second bin of a video session. As
with our startup delay analysis, we study the correlation between resolution
and access capacity. For this analysis, we omit the first 60~seconds of video
sessions, since many Adaptive Bitrate (ABR) algorithms begin with a low
resolution to ramp up as they obtain a better estimate of the end-to-end
throughput. Recall that our video quality inference model outputs one of five
resolution classes: 240, 360, 480, 720, or 1080p.

Figure~\ref{fig:nominal} presents resolution versus the nominal speed tier,
whereas Figure~\ref{fig:bismark} presents resolution versus the 95th percentile
of active throughput measurements. As with startup delay, we observe expected
results (\ie, higher resolutions with higher capacities) with measured
throughput (Figure~\ref{fig:bismark}); the trend is less clear for nominal speed
tiers (Figure~\ref{fig:nominal}). Recall that the number of samples in the
highest speed bar for Amazon and Twitch is small; focusing on the other three
tiers shows a clear trend of increased percentage of bins with higher
resolutions as capacity increases. For example, Netflix and YouTube in the
highest speed tier achieve about 40\% more 1080p than in the lowest speed tier.
We also observe that, in general, YouTube streams at lower resolution for the
same network conditions than other services.

Video resolution is dependent on more factors than simply the network conditions.
First, some videos are only available in SD, and thus, will stream at 480p
regardless of network conditions. Second, services tailor to the device type as
a higher resolution may not be playable on all devices.
We can identify the device type for some of the devices in the deployment based
on the MAC address. Of 1,290,130 ten-second bins of YouTube data for which we
have device types, 616,042 are associated with smartphones, indicating that
device type may be a confounder.

\balance\section{Conclusion}\label{sec:conclusion}

Internet service providers increasingly need ways to infer the quality of
video streams from encrypted traffic, a process that involves both identifying
the video sessions and segments and processing the resulting traffic to infer
quality across a range of services. We build on previous work that infers
video quality for specific services or in controlled settings, extending the
state of the art in several ways.  First, we infer startup delay and
resolution delay more precisely, attempting to infer the specific values of
these metrics instead of coarse-grained indicators. Second, we design a model
that is robust in a deployment setting, applying techniques such as domain
adaptation to make the trained models more robust to the noise that appears in
deployment, and developing new techniques to identify video sessions and
segments among a mix of traffic that occurs in deployment.  Third, we develop
a composite model that can predict quality for a range of services: YouTube,
Netflix, Amazon, and Twitch.  Our results show that prediction models that use
a combination of network- and application-layer features outperforms models
that rely on network- and transport-layer features, for both startup delay and
resolution. Models of startup delay achieve less than one second error for
most video sessions; the average precision of resolution models is above 0.93.
As part of our work, we generated a comprehensive labeled set with more than
13,000 video sessions for four popular video services: YouTube, Netflix,
Amazon, and Twitch.  Finally, we applied these models to 16 months of traffic
from 66 homes to demonstrate the applicability of our model in practice, and
to study the relationship between access link capacity and video
quality. We found, surprisingly, that higher access speeds provide only
 marginal improvements to video quality, especially at higher speeds.

Our work points to several avenues for future work. First, our composite model
performs poorly for services that are not in the training set; a truly general
model that can predict video quality for arbitrary services remains an open
problem.  Second, more work remains to operationalize video quality prediction
models to operate in real time.
\label{p:lastpage}\clearpage
\end{sloppypar}

\small
\setlength{\parskip}{-1pt}
\setlength{\itemsep}{-1pt}
% \footnotesize % SPACE
\balance\bibliography{bibliography}
\bibliographystyle{ACM-Reference-Format}
%\bibliographystyle{abbrvnat_noaddr} % SPACE
%\theendnotes % ENDNOTES

\end{document}